\documentclass[letterpaper,twocolumn,10pt,hyphens]{article}
\usepackage{usenix2019_v3}

\microtypecontext{spacing=nonfrench} 

\newif\ifblind
\blindfalse

\usepackage[suppress]{color-edits}
\addauthor{aa}{red}
\addauthor{bk}{blue}
\addauthor{bw}{cyan}
\addauthor{sr}{magenta}

\usepackage{inconsolata}   
\usepackage{tikz,ifthen}
\usetikzlibrary{math}
\usepackage{amsmath,amsthm,amsfonts,amssymb,bm,bbm,mathrsfs} 
\usepackage{xspace}
\usepackage{hyperref}
\hypersetup{colorlinks=true,linkcolor=blue}
\usepackage[nameinlink]{cleveref}
\usepackage{enumitem}
\usepackage[labelfont=bf]{caption}
\usepackage{subcaption}
\usepackage{titlesec}
\usepackage{csquotes}
\usepackage{xcolor}
\usepackage{listings}
\usepackage{soul}

\theoremstyle{definition}

\crefformat{section}{\S#2#1#3}
\crefmultiformat{section}{\S\S#2#1#3}{ and~#2#1#3}{, #2#1#3}{, and~#2#1#3}
\Crefformat{section}{\S#2#1#3}
\Crefmultiformat{section}{\S\S#2#1#3}{ and~#2#1#3}{, #2#1#3}{, and~#2#1#3}
\crefname{figure}{fig.}{figs.}

\titlespacing{\paragraph}{%
  0pt}{
  0.2\baselineskip}{
  1em}

\newcommand{\half}{\ensuremath{\frac{1}{2}}}

\newcommand{\ie}{i.e.,\xspace}
\newcommand{\eg}{e.g.,\xspace}

\newcommand{\hotcold}{HCL\xspace} 
\newcommand{\plb}{Prequal\xspace}
\newcommand{\plbexp}{Probing to Reduce Queuing and Latency\xspace}
\newcommand{\podc}{PodC\xspace} 
\newcommand{\rprobe}{r_{\text{probe}}}
\newcommand{\rremove}{r_{\text{remove}}}
\newcommand{\rifpercent}{\ensuremath{Q_{\text{RIF}}}\xspace}
\newcommand{\riflimit}{\ensuremath{\theta_{\text{RIF}}}\xspace}
\newcommand{\removeworst}{\ensuremath{\rremove}\xspace}
\newcommand{\reuselimit}{\ensuremath{b_{\text{reuse}}}\xspace}
\newcommand{\qps}{QPS\xspace}
\newcommand{\roughly}{\textasciitilde}
\newcommand{\wrr}{WRR\xspace} 

\definecolor{pink}{rgb}{1,0.90,0.95}
\sethlcolor{pink}
\definecolor{maroon}{rgb}{0.6,0,0}
\newcommand{\resub}[1]{\xspace{#1}\xspace}

\begin{document}

\date{}

\title{\Large \bf Load is not what you should balance: Introducing \plb}

\ifblind
\author{\large \itshape Operational Systems Track}
\else
\author{
{\rm Bartek Wydrowski}\\
Google Research
\and
{\rm Robert Kleinberg}\\
Google Research
\and 
{\rm Stephen M. Rumble}\\
Google (YouTube)
\and
{\rm Aaron Archer}\\
Google Research
}
\fi

\maketitle

\begin{abstract}
We present Prequal (\emph{Probing to Reduce Queuing and Latency}), a load balancer 
for distributed multi-tenant systems. Prequal aims to minimize 
real-time request latency in the presence of heterogeneous server 
capacities and non-uniform, time-varying antagonist load. It actively probes 
server load to leverage the \emph{power of $d$ choices} 
paradigm, extending it with asynchronous and reusable probes. Cutting 
against received wisdom, Prequal does not balance CPU load, but instead 
selects servers according to estimated latency and active requests-in-flight 
(RIF). We explore its major design features on a testbed system 
and evaluate it on YouTube, where 
it has been deployed for more than two years. Prequal has dramatically 
decreased tail latency, error rates, and resource use, enabling YouTube and
other production systems at Google to run at much higher utilization.

\end{abstract}

\section{Introduction}
\label{sec:intro}


\resub{
We report our experience deploying the power of $d$ choices (\podc) load 
balancing paradigm \cite{azar99balanced,Mitzenmacher96-phdthesis,Kanellakis-Award-2020-power-of-two-choices}
to run multiple large-scale web services at Google, 
focusing on YouTube, where it has run successfully for two years. To our 
knowledge, this is the first public report of \podc being used successfully \bkedit{for load balancing} at \bkedit{this} scale. 

The \podc paradigm 
involves sampling $d \geq 2$ servers for their load, and sending the next 
request to the server with minimum load. The two central questions for any 
implementation are: (1) what signal is used to represent the load, and (2) 
how is the sampling done? We answer those questions with the name of our 
load balancing system: Probing to Reduce Queuing and Latency (\plb). Namely, 
we use two signals: requests-in-flight (RIF) and latency, and sample servers by 
actively probing them.

Many existing load balancing systems 
(\eg NGINX \cite{nginx-load-balancing}, Envoy \cite{envoy-homepage}, Finagle \cite{finagle-po2c}, 
YARP \cite{YARP-homepage}, C3 \cite{suresh15c3}) 
offer some variant of \podc, 
and most of them offer RIF, latency, or some combination as the load 
balancing signal. 
We introduce two primary new innovations. First, we combine 
RIF and latency in a new way that works especially well, called the \emph{hot-cold lexicographic} 
(\hotcold) rule. Second, we introduce a novel asynchronous probing mechanism that 
reduces probing overheads (CPU + critical-path latency) while retaining 
the freshness of the load signal offered by synchronous probing.

For purposes of this paper, we refer to \plb as a load balancing \emph{system}, 
but technically, it is load a balancing \emph{policy} implemented within Google's 
Stubby RPC framework (externally, gRPC \cite{grpc-blog}). As such, we expect that some of our innovations could be 
integrated into these other existing load balancers. Our main thrust is the two innovations 
above (\hotcold, async probing); other implementation details of our system are incidental to our message.

\plb has been deployed in a diverse collection of 20+ large-scale services at Google over the past 
two years. In addition to 
driving most of the YouTube serving stack, it has found success in a wide variety of other applications, 
including search ads, logs processing, and serving ML models. In the typical application, query processing 
times are $O(10-100)$ milliseconds, but in one system the queries take
$O(10)$ minutes, and in another they range from seconds to hours. Most of these services are part of a 
complex tangle of services calling each other, and we have seen benefits from deploying \plb for the 
most critical services, regardless of whether others in the tangle are also using \plb. 

We care about CPU and latency overheads from probing, but in our deployment experience we have found them to be
small and more than pay for themselves. Probe response times within a data center are well 
below 1 millisecond. For some applications, 
\bkedit{fractions of a millisecond matter}, so we used async probing to take it off the critical path. We 
created parameters that can shrink the CPU overheads arbitrarily. Moreover, better balance saves CPU by reducing resource contention. On net, we find that CPU utilization usually \emph{dips} slightly when deploying \plb. More importantly, systems are provisioned to control tail latency at peak load, so even an \emph{increase} in \emph{mean} CPU usage can result in provisioning wins when combined with decreases in tail latency.

Our evaluation of \plb consists of two parts: a holistic evaluation in the 
wild on YouTube (\Cref{sec:motivation} and \Cref{sec:youtube}), and benchmarks on a testbed to evaluate each 
of the major design choices independently (\Cref{sec:evaluation}). The default incumbent 
load balancing policy that we displaced at YouTube and other parts of Google is 
(dynamic) weighted round robin (\wrr), which focuses on balancing CPU utilization 
across distributed servers in a single job. Thus, the evaluation in \Cref{sec:youtube} 
focuses on comparing \plb to \wrr.  
\Cref{sec:motivation} explains why CPU-balancing policies like \wrr are a formidable competitor in this 
environment, but also why it pays to 
rethink that approach, motivating our paper title. \Cref{sec:evaluation} and \Cref{sec:related} include 
quantitative and qualitative comparisons 
against well-known load balancing policies from the literature, including other variants of \podc. 
}

\aadelete{A large-scale web service such as YouTube
comprises many subsidiary jobs --- answering
search queries, providing recommendations, 
and serving ads, to name a few --- 
each processing a multitude of queries 
every second. Managing datacenter
resources to run such services efficiently 
and reliably is a major challenge. To avoid
excessive overprovisioning, systems need flexibility 
to allow some jobs to momentarily exceed their allocated 
resource limit when others running on the same machine are underutilized. 
This is at odds with 
performance isolation between jobs, endangering 
the system's ability to reliably deliver 
low-latency responses to every query.}
\aadelete{In this paper we argue that the tension 
between provisioning efficiency and serving reliability requires
rethinking the paradigm for load balancing
in systems -- such as large-scale web services -- 
where multiple clients each governed by their 
own load balancer share the same pool of 
servers. In particular, the aim of a load 
balancer in such systems should not be to distribute
its client's load evenly among the available 
servers. The ultimate goal is to deliver reliable low latency at high throughput,
so load balancers should aim to minimize
tail latency and be evaluated using metrics that 
align with this goal.}
\aadelete{In this work, we introduce 
\emph{\plbexp} (\plb), a load balancer 
based on actively probing available servers to 
monitor their latency and number of requests 
in flight (RIF). \plb uses the \emph{power of $d$ choices} 
paradigm \cite{azar99balanced,Mitzenmacher96-phdthesis,Kanellakis-Award-2020-power-of-two-choices}
to avoid selecting overloaded servers. 
It enhances the basic paradigm with asynchronous and 
reusable probes, as well as a customizable trade-off 
between favoring servers with low latency and those 
with few requests in flight.}



\section{\resub{Environment and} motivation}
\label{sec:motivation}

\resub{
We describe the operating environment for YouTube and other large cloud services, 
motivating why balancing CPU load would seem to be a self-evidently superior approach.
\aacomment{I previously wrote "seems to be a no-brainer," but decided that was too informal.}
\aadelete{Maybe this is valuable, but I deleted it for space:
Indeed, \wrr does outperform many other popular load balancing 
policies in this environment, which is why it was the main incumbent solution at YouTube.}
Next, we examine a scenario in which balancing CPU load backfires spectacularly, 
motivating the \plb approach. Finally, we show data from YouTube 
illustrating that our motivating circumstance occurs commonly.}

\begin{figure}
    \centering
    \includegraphics[width=0.45\textwidth,keepaspectratio]{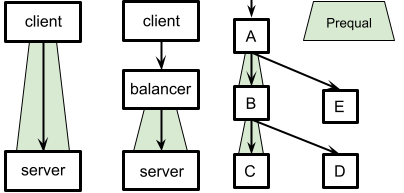}
    \caption{Queries flow down the tree, responses flow up. A separate balancing job is optional on each link. Load balancing policies may vary by link: A $\rightarrow$ B uses \plb, while B $\rightarrow$ D does not.}
    \label{fig:query-flow}
\end{figure}

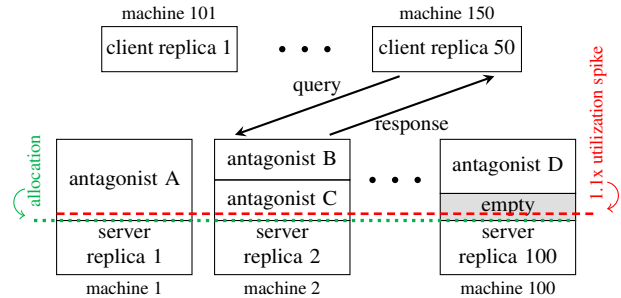
\begin{figure}[!t]
    \centering
    \begin{tikzpicture}[scale=0.6,every node/.style={font=\footnotesize}]
        \draw (1.5,5.8) node {\scriptsize machine 101};
        \draw (0.0,4.5)
        rectangle (3.0,5.5)
        node[midway] {client replica 1};
        \draw (7.65, 5.8) node {\scriptsize machine 150};
        \draw (6.0,4.5)
        rectangle (9.3,5.5) node[midway] {client replica 50};
        \draw[-stealth, line width=0.8] (6.6,4.4) -- node[above] {query} (2.95,3.1);
        \draw[stealth-, line width=0.8] (8.7,4.4) -- node[below=2pt] {response} (5.05,3.1);
        \fill (4.0,5.0) circle (2pt);
        \fill (4.5,5.0) circle (2pt);
        \fill (5.0,5.0) circle (2pt);
        \draw (-1.0,0.0) rectangle (2.0,1.2)
        node[midway,text width=35pt,align=center] {server replica 1};
        \draw (-1.0,1.2) rectangle (2.0,3.0) node[midway,text width=45pt,align=center] {antagonist A};
        \draw (2.5,0.0) rectangle (5.5,1.2) node[midway,text width=45pt,align=center] {server replica 2};
        \draw (2.5,1.2) rectangle (5.5,2.1) node[midway,text width=45pt,align=center] {antagonist C};
        \draw (2.5,2.1) rectangle (5.5,3.0) node[midway,text width=45pt,align=center] {antagonist B};
        \fill (6.0,2.1) circle (2pt);
        \fill (6.5,2.1) circle (2pt);
        \fill (7.0,2.1) circle (2pt);
        \draw (7.5,0.0) rectangle (10.5,1.2)
        node[midway,text width=45pt,align=center] {server replica 100};
        \fill[gray!25] (7.5,1.2) rectangle (10.5,1.8);
        \draw (7.5,1.2) rectangle (10.5,1.8)
        node[midway,text width=45pt,align=center] {empty};
        \draw (7.5,1.8) rectangle (10.5,3.0)
        node[midway,text width=45pt,align=center] {antagonist D};
        \draw[dotted, line width=1.25,blue!30!green] (-1.5,1.2) -- (10.5,1.2);
        \draw[densely dashed, line width=1,red]
        (-1.0,1.35) -- (11,1.35);
        \draw (-1.5,2.3) node[blue!30!green,rotate=90] {\scriptsize allocation};
        \draw[->,blue!30!green] (-1.7,1.9) arc [
            start angle=100,
            end angle=260,
            x radius=0.3,
            y radius=0.3
        ];
        \draw (11,3.35) node[red,rotate=90] {\scriptsize 1.1x utilization spike};
        \draw[->,red] (11.2,2.05) arc [
            start angle=80,
            end angle=-80,
            x radius=0.3,
            y radius=0.3
        ];
        \draw (0.5,-0.3) node {\scriptsize machine 1};
        \draw (4.0,-0.3) node {\scriptsize machine 2};
        \draw (9.0,-0.3) node {\scriptsize machine 100};
    \end{tikzpicture}
    \caption{\resub{Zooming in on any single link from \Cref{fig:query-flow}:} server replicas respond to queries from client replicas while their machines run antagonist processes. Aggregate utilization (dashed line) may exceed allocation (dotted line) during spikes in demand.}
    \label{fig:system-diagram}
\end{figure}

\resub{
Large-scale services like YouTube are typically composed of many distributed 
jobs issuing queries among themselves via RPC. \Cref{fig:query-flow} 
depicts a 
vastly simplified version with only five jobs, queries flowing down a tree (with the arrows), 
and responses flowing back up (against the arrows). In any one of these interactions, the job 
sending the query is called the \emph{client job}, while the job returning 
the response is the \emph{server job}. For instance, on edge A $\rightarrow$ B, 
A is the client and B is the server, while B is the client on edges B $\rightarrow$ C and B $\rightarrow$ D. 
To ensure scalability and redundancy, each job is distributed across 
a large number of machines, called \emph{replicas} (\Cref{fig:system-diagram}). To avoid 
overloading any server replica, a load balancing policy must be used between 
each pair of communicating jobs, to determine which server replica 
should receive each query. Some policies require proxying the queries 
through a separate load balancing job whose only function is to decide where 
to forward each query, while others can operate directly 
between client replicas and server replicas (\Cref{fig:query-flow}).\footnote{At this scale, even the balancing job must be distributed across machines.} 
\plb \bkedit{is well-suited to both modes of operation}, and in fact we use it both ways in 
different systems at Google.

Each job runs within a single data center, although different jobs might 
run in different data centers. Each replica runs inside its own virtual machine (VM), which}
has access to a guaranteed portion of the CPU cycles on its host 
machine; this amount is called its \emph{(CPU) allocation}.
The fraction of its 
allocation that a replica uses over any 
given time period is 
its \emph{(CPU) utilization}, which could be more or less than 100\%, since the 
allocation is just a 
guaranteed minimum. The server job's \emph{aggregate (job) utilization} is 
the fraction of 
its overall CPU allocation that it uses, across all replicas.
In the usual case, each replica has the same allocation, 
so aggregate \resub{job} utilization equals average \resub{replica} utilization.

\resub{A server replica typically shares its machine with many other VMs, which we call \emph{antagonists}.} 
The \emph{machine (CPU) utilization} accounts 
for both our server replica and all antagonists, and is distinct from both 
replica and job utilizations.
\resub{In order to ensure consistent performance for all users 
of this multi-tenant system, \emph{isolation} mechanisms attempt 
to prevent the behavior of one 
VM on a machine from adversely affecting the behavior of another. The 
philosophy here is: "If your usage stays within your allocation, you will 
be fine." However, if a VM overflows its CPU allocation, it may suffer when the 
isolation system kicks in.}

\resub{
Given this isolation regime and design philosophy, it \emph{seems} self-evident that trying to balance CPU \bkedit{utilization} across replicas within one job is the best approach.\footnote{\resub{Trying to balance \bkedit{machine CPU utilization} across these entire heterogeneous machines does not make sense.
}} Indeed, the \wrr policy works well at Google.
It uses smoothed historical statistics on each replica's goodput, CPU utilization, and error rate to periodically compute
individual per-replica weights. Clients then route queries to replicas in
proportion to these weights. In the absence of errors, each replica weight $w_i$ is
calculated as $q_i/u_i$, where $q_i$ and $u_i$ represent the recent query-per-second (\qps) rate 
and CPU utilization of replica $i$. Historically, WRR originated in networking~\cite{katevenis1991weighted} 
but has also been adapted for replica selection as we just 
described~\cite{Google-SRE-book-load-balancing, gRPC-software-WRR}.

We now describe a scenario where this approach backfires, even if we could balance CPU load \emph{perfectly}.} Suppose that our server job has 100 replicas running on identical machines, each replica 
allocated 40\% of the CPU on its machine. Moreover, suppose that 
antagonist load is soaking up the full remaining 60\% on machines 1 and 2, whereas the other 98 
machines have ample spare capacity, safely above 4\%. Finally, suppose that our job experiences 
a temporary spike in demand that would raise its aggregate utilization to 1.1x its allocation, \ie 44\% of each machine, on average.
A load balancer that attempts to equalize 
CPU utilization across replicas will aim to peg each replica at this average.
All 100 replicas will exceed their CPU allocation, but the last 98 will be okay because the system will let them momentarily spill 
outside their allocation to soak up the unused CPU cycles. On the heavily contended machines 1 and 2, 
CPU isolation mechanisms will typically kick in and hobble those replicas, sometimes in ways that affect \emph{all} queries served by them.
Thus, although the problematic load is only the extra 10\% on only 2\% of our replicas (\ie \roughly 0.18\% of our 1.1x load), 
it could degrade performance on a full 2\% of our queries. Tail latency at p99\footnote{We use expressions like p99 to denote percentiles of a distribution.} is likely to spike. 
\Cref{fig:load-ramp} in \Cref{sec:wrr} exhibits this behavior on a testbed. 


\begin{figure}[!t]
    \centering
    \includegraphics[width=0.45\textwidth,keepaspectratio]{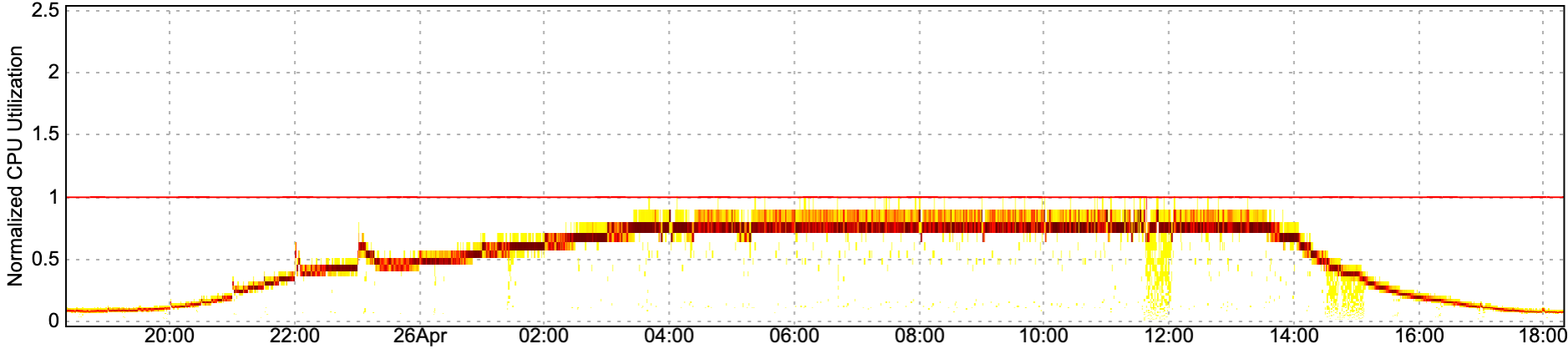}
    \includegraphics[width=0.45\textwidth,keepaspectratio]{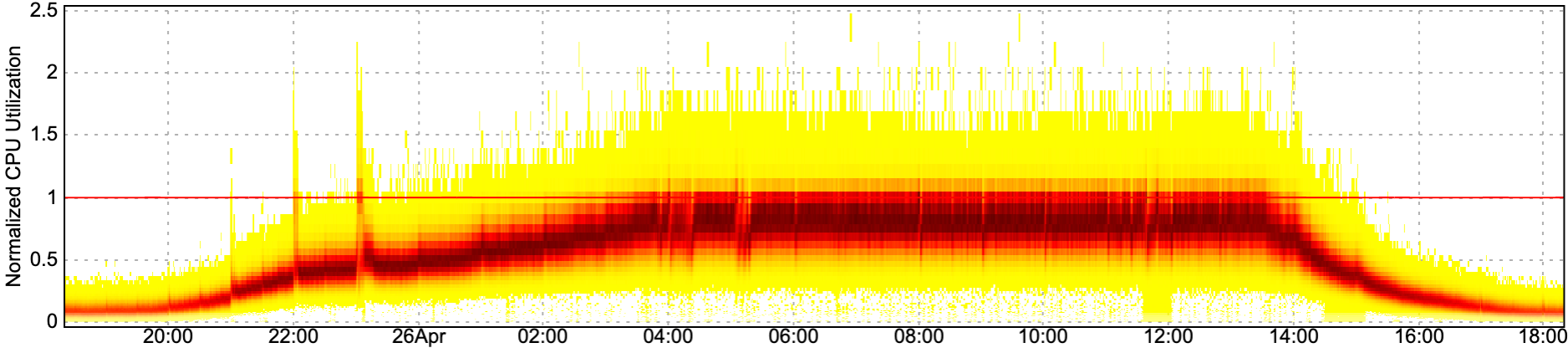}
    \caption{Normalized CPU usage heatmap with 1-minute samples
    (top) and 1-second samples (bottom) for the YouTube homepage service running across hundreds of machines in a single datacenter. Darker colors represent denser regions. The usage
    limit, represented by 1.0 (red line) is satisfied in all 
    of the 1m intervals but on 1s intervals it is frequently violated at peak load
    --- sometimes by more than a factor of two!}
    \label{fig:youtube-timescale-cpu}
\end{figure}


In this overload case, equalizing replica utilizations was exactly the 
wrong load-balancing policy because the
available machines differed greatly in 
their capacity to absorb additional load. 
Furthermore, since the difference in 
available capacity is due to antagonist
processes, it cannot be predicted by our job in 
advance but can potentially be detected
at runtime.


Equalizing replica utilizations can be a great idea if all replicas always stay within 
their allocation, as this maximizes the 
amount of traffic we can serve, subject to the allocation. Unfortunately,
\Cref{fig:youtube-timescale-cpu} shows it is easy to trick ourselves. Plotting 
CPU utilization over 1m time intervals leads us to believe that the replicas are all
respecting their allocations, but using 1s intervals reveals
greater underlying variability in the signal, with frequent bursts up to nearly twice the limit!
In other words, overload is not really a special case; at sufficiently small timescales, there is nearly always \emph{some} replica in overload, 
even if our aggregate load fits within our job allocation. 
The only questions are whether this replica is unlucky enough to be on a highly contended machine, and whether the spike 
lasts long enough for isolation mechanisms to kick in.


Thus, avoiding high tail latency requires some
mechanism to alert clients quickly about highly-loaded
replicas in real-time. The mechanism should use load
signals that are as current as possible, 
and are highly predictive of high 
latency when serving future requests.
CPU utilization fails to meet these criteria, partly because it 
is a trailing signal. It must be averaged 
over a time window to be meaningful, which 
automatically 
imposes a lower bound on its staleness. 
This signal also overlooks other factors that contribute
to latency, like contention for locks, memory 
bandwidth, or other shared resources that are 
sometimes tough to measure or isolate.

\plb instead uses two load signals:
RIF and latency. RIF is an instantaneous
signal (\ie its precise value is available 
at the time of a probe), and the latency
estimates used by \plb are near-instantaneous
as well (\Cref{sec:design}). Here are our main design goals.
\begin{enumerate}
    \item \aadelete{The probing process should place
    minimal resource demands on replicas.} \resub{Minimize probing overheads.
    The number of probes per query should be a small $O(1)$}
    and the latency estimation algorithm (running on server replicas) 
    must be lightweight, taking
    \aaedit{$O(1)$ or $\Tilde{O}(1)$} update time per query.
    \item \aadelete{Sending probes and
    waiting for results should not add 
    significantly to the latency for
    processing queries.} \resub{Probing should not add significant latency 
    to the query's critical path.} \plb accomplishes
    this via asynchronous probing: 
    the current query is assigned using probe
    responses initiated by previous queries.
    \item \resub{Minimize tail latency.} 
    \plb removes the worst 
    probes (those with the highest RIF and/or estimated 
    latency) before they can be used for 
    replica selection, in a process inspired 
    by the theory of balanced allocations
    with memory~\cite{los23memory,mitzenmacher02memory}.
    \item \aadelete{In addition to avoiding high tail 
    latency, }
    \resub{Limit RAM footprint of query processing on server replicas.}\footnote{\resub{RIF has a material impact on RAM because most in-flight queries are 
    in some stage of processing, \emph{not} sitting in a queue. The RPCs themselves are often  
    small.}} 
    \plb avoids assigning queries to replicas that have an
    anomalously large RIF, even if their estimated
    latency is low. The RIF signal does double duty, since it is 
    also a strong leading indicator of future load.
\end{enumerate}

\aadelete{
Note that \plb is a completely distributed load balancing solution. 
Client replicas communicate only with server replicas, so they do not 
share state among themselves.
In general, there is no dedicated load balancing
layer sitting between the client and server replicas. The only exception is 
when the source of the queries sits far enough away from the server replicas that 
the lightpath delay is significant, thereby causing the probing delay to pose a problem.
In this case, we can run a balancer job consisting of local \plb client replicas colocated 
near the server replicas. 
The remote client sends the query to a local client replica using any fallback policy, and the local client
runs \plb, performing probing locally. In our experience with YouTube, the 
resources required by the
local balancer job are minuscule relative 
to the server replicas 
and any additional costs are greatly exceeded by savings obtained from improved load balance 
and tighter provisioning. For the same reason, load balance on the balancer job itself is a lesser concern,
since we can afford to overprovision it. For the remainder of the paper, we consider only the case where 
clients and server replicas are colocated.}

\resub{
Recall that \plb can be used either with or without a dedicated load 
balancing layer (\Cref{fig:query-flow}). 
Advantages of the dedicated layer include (1) keeping probes local 
when clients query server jobs in a distant 
data center, (2) the balancer often has fewer replicas than the 
client does, so each one sees a larger fraction of the query stream, 
hence its probes are fresher (as measured by number of queries landing on a 
server replica since the most recent probe), (3) software upgrades to the 
balancer need not touch the clients.
Disadvantages include (1) 
further RPC overhead (latency, CPU, network), (2) it is an extra job to 
manage, and (3) 
one still needs to balance the load arriving at the balancers, although that 
is usually easier because the balancer's work is fairly uniform, and it is 
usually cheaper to overprovision the balancer than the server.
}



\aadelete{\section{Deployment experience in YouTube}}
\section{\resub{Evaluation of deployment in YouTube}}
\label{sec:youtube}


A couple of years ago, we identified load imbalance as the cause of persistent
SLO\footnote{SLO = service level objective~\cite{Google-SRE-SLO}} violations \bkedit{in YouTube}, prompting us to switch some key services to \plb. These included YouTube's Homepage service, which is responsible for user-tailored recommendations. With \plb we saw dramatic improvements across all metrics, 
including reductions of 2x in tail latency, 5-10x in tail RIF, 10-20\% in tail memory usage, 2x in tail CPU utilization, and a near-elimination of errors due to load imbalance. In addition to meeting SLOs, this also allowed us to significantly raise the utilization targets that govern how much traffic we are willing to send to each datacenter, thereby saving significant resources. 
Based on this success, we rolled out \plb to many other major services that 
comprise YouTube, and 
\bkedit{subsequently to a wide variety of other services at Google, as detailed in \Cref{sec:intro}.}

\resub{The YouTube service is composed of many subsidiary jobs --- answering 
search queries, providing recommendations, and serving ads, to name a few. 
Queries from some of these jobs trigger a cascade \bkedit{of subsidiary queries}, as illustrated in 
\Cref{fig:query-flow}. Since these jobs are so diverse, their query 
processing requirements (CPU, RAM, latency, etc.) vary widely as well.
In this section we show the results of switching from \wrr to \plb using a small subset of live, user-facing traffic
for the Homepage job itself, with the load 
balancing policies in other parts of the system held constant. Historically, the 
Homepage was the first server job where we deployed \plb, while most other jobs still 
used \wrr. At the time of this experiment, most but not all of these were using \plb. 
In both cases, the impact of the switchover was similar.}
%
%
%
We state \plb's impact here, but defer the details of its design to \Cref{sec:design}. We then explore this 
design space via experiments on a testbed (not YouTube) in \Cref{sec:evaluation}.

\begin{figure}
    \centering
    \includegraphics[width=0.45\textwidth,keepaspectratio]{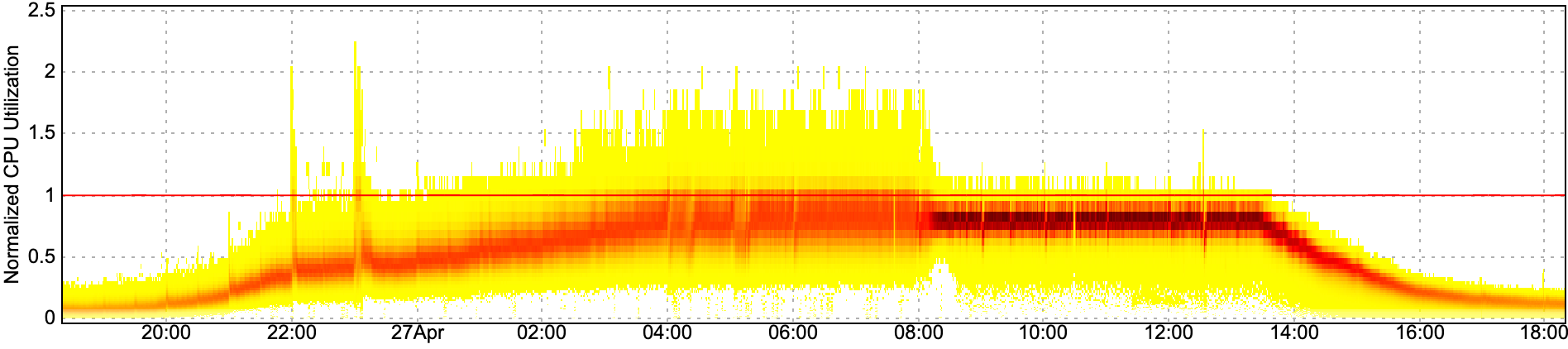}
    \includegraphics[width=0.45\textwidth,keepaspectratio]{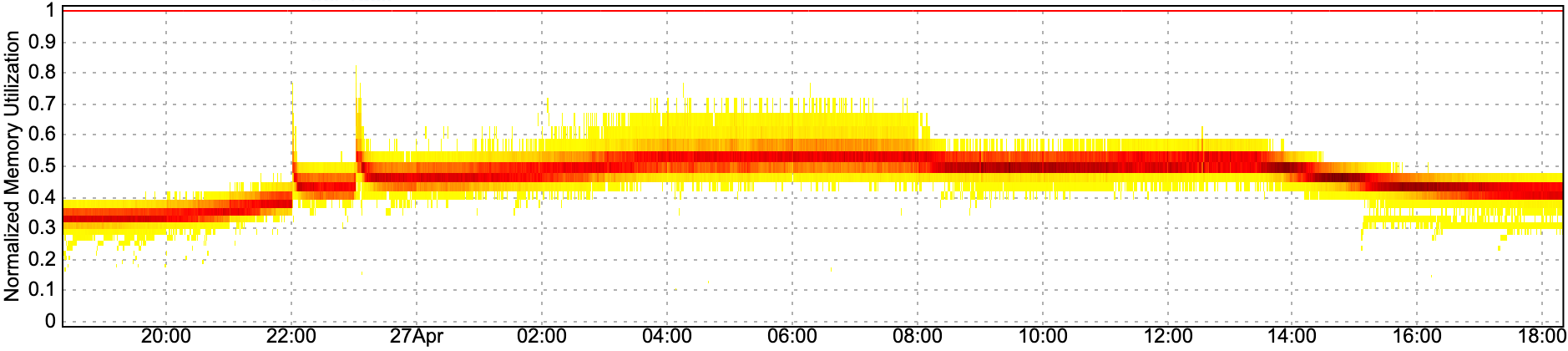}
    \includegraphics[width=0.45\textwidth,keepaspectratio]{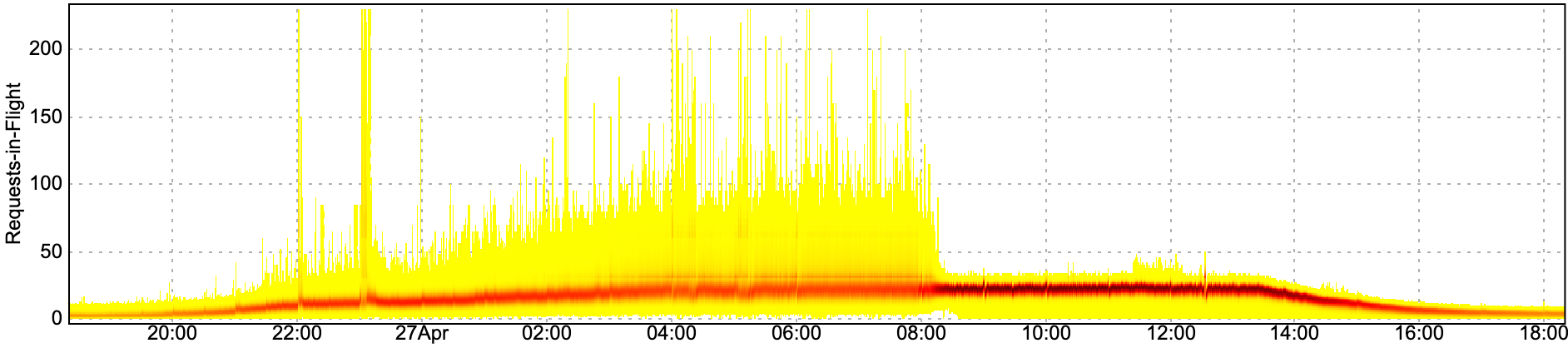}    
    \caption{Heatmaps of normalized cpu usage, normalized memory usage, and the number of requests in flight on each YouTube Homepage server replica, first using WRR to balance load before transitioning to \plb shortly after 08:00.}
    \label{fig:youtube-rif}
\end{figure}

\begin{figure}
    \centering
    \includegraphics[width=0.45\textwidth,keepaspectratio]{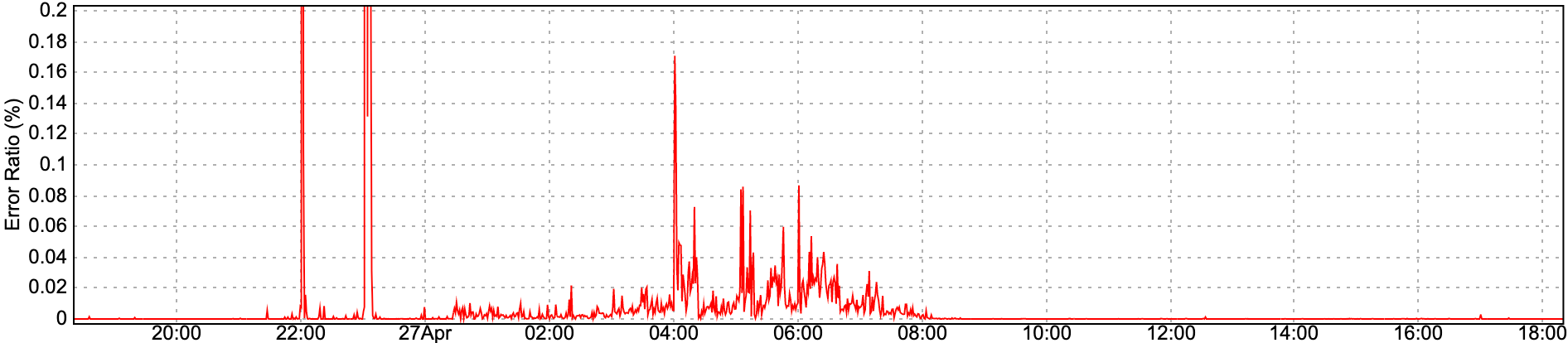}
    \includegraphics[width=0.45\textwidth,keepaspectratio]{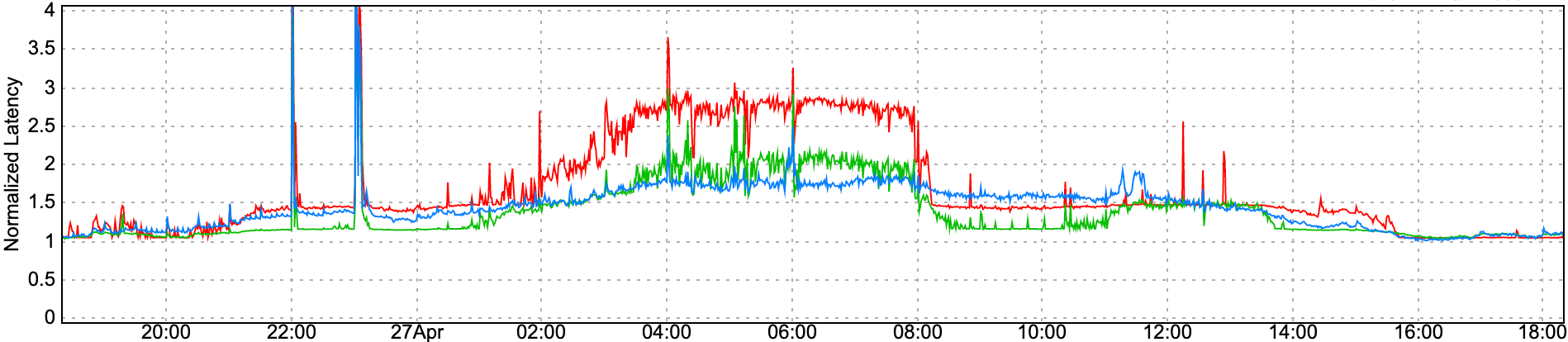}
    \caption{Normalized YouTube Homepage server request error rate and latency p99.9 (red), p99 (green), and p50 (blue). Failures 
    were due to timeouts or load shedding stemming from imbalance. Cutting over from WRR to \plb shortly after 08:00 eliminated 
    most errors and reduced tail latency by 40-50\% and median latency by 5-10\%. Each latency quantile is 
    normalized separately to a typical value at daily trough \emph{for that quantile}, which is why it is possible for the p50 line to lie
    above the other two after the cutover. The spikes near 22:00 and 23:00 are unrelated to the experiment.}
    \label{fig:youtube-errors-latency}
\end{figure}

The first result to notice is that explicitly balancing on RIF really 
works, bringing down the tail from \roughly 225 to \roughly 50 (\Cref{fig:youtube-rif}). Since Homepage query 
processing carries a large amount of per-query state, this reduces our tail RAM usage 
by 10-20\%, allowing us to reduce our RAM footprint 
accordingly.
As shown in \Cref{fig:youtube-timescale-cpu}, WRR was very effective 
at maintaining a tight CPU load distribution at the 1m time scale, but
much poorer when measuring tail CPU utilization at 1s resolution. \plb fixes that, 
dropping the tail by \roughly 2x. As a consequence, \plb decreased occasional server replica error spikes of 0.01-0.1\% down to nearly zero and reduced latency by 10-20\% at the median and 40-50\% at the tail (\Cref{fig:youtube-errors-latency}).

\Cref{fig:youtube-errors-latency} normalizes each latency quantile (p50, p99, p99.9) 
separately, based on a typical value for 
\emph{that quantile} at the daily traffic trough. This reveals that \plb does such a 
good job of pulling in the tail latency at peak load that 
the p99 and p99.9 actually suffers \emph{less} at peak (in a multiplicative sense) than p50 does. This is the opposite of the 
behavior one would normally expect, and that we indeed see for WRR. 


For these experiments, we configured \plb to send 5 probes per query, which means that the total number of 
RPCs is multiplied by 6. Moreover, as an early adopter of \plb, this service still uses the synchronous 
probing mode (\Cref{sec:design}), which adds latency to the critical path. While we intend to migrate to asynchronous probing, 
our experience with YouTube shows that the improvements we get by pulling in the tails more than compensates 
for these overheads. This is particularly true for services with relatively heavyweight requests, \eg 
\aaedit{$O(10-100)$ milliseconds}
of processing time per query. We set a 3ms timeout for our probes, but most return far quicker 
than that.\footnote{Elsewhere at Google, we successfully use a 1ms timeout, and could probably do so here too.} The probing delay and the extra CPU spent on processing the probes are both in the noise compared 
to the savings.

\section{System design}
\label{sec:design}

When load-balancing a torrent of queries, every millisecond matters. \plb is designed so that clients can select a replica based on up-to-date information (ideally no more than a few \bkedit{milliseconds} old) while keeping the process of sending and receiving probes out of the critical serving path. Hence, the most consequential design decisions center around how to maintain a pool of high-quality probes and how to assign queries to \bkedit{replicas} based on the information in the pool. 
\bkdelete{In the broadest outline, a client running \plb probes replicas for load information, maintains a pool of probe responses, and uses the information in that pool to assign new queries to replicas. Within this broad outline, there are several key design questions.} 
\bkedit{The following are some key questions.}
\begin{enumerate}[itemsep=1pt,topsep=3pt,partopsep=0pt]
    \item How often should a client probe replicas, and which ones should it probe?
    \item What information should \bkedit{replicas} report in probe responses, and how should they compute it?
    \item When selecting a replica to handle a query, how should a client make use of the information in its probe pool?
    \item How should a client manage its pool of probe responses? In particular, when should elements be removed\bkdelete{ from the pool}?
\end{enumerate}
\paragraph{Probing rate}
Clients issue a specified number of probes\bkedit{,
$\rprobe$,} triggered by each query. 
In addition, they can be configured to issue probes 
after a specified maximum idle time has been exceeded, 
to ensure the availability of recent probe responses
in the pool even when no queries have
arrived in the recent past. The probing rate (per
unit time) is therefore the product of the query arrival
rate with \bkedit{$\rprobe$} (or the minimum
probing rate, whichever is greater). We link the probing 
rate to the query rate because that is the rate at which 
the client must make decisions. It is also strongly tied 
to how rapidly the load statistics on the replicas (particularly 
RIF) change over time, although that also depends on the ratio of 
client replicas to server replicas.
Probing at a rate that greatly exceeds the query 
arrival rate would thus yield \bkedit{many}
redundant probes, whereas probing at \bkedit{too low} a rate 
runs the risk
that probe results will be ``stale'' (only weakly 
correlated with the replica's current state) by the
time they are used. 

Probe destinations are sampled uniformly at random 
without replacement from the set of available replicas. 
This design choice is motivated by the theory literature 
on balanced allocations, which advocates sampling a 
uniformly random set of probe targets. It also helps
to avoid the \emph{thundering herd} phenomenon, when 
\bkedit{a replica with low estimated latency 
is inundated with queries from many clients 
simultaneously seeing it as the best choice,
leading to queueing and high latency~\cite{mitz00useful}.
This is unlikely to happen in \plb, where each client's
probe pool represents only a small random subset of replicas,
and conversely, each replica belongs to the probe pools of
only a small (in expectation) random subset of clients.}

We allow $\rprobe$ to be 
fractional
\footnote{
Each query triggers either $\lfloor \rprobe \rfloor$ or $\lceil \rprobe \rceil$ probes, rounding deterministically 
so as to guarantee $\rprobe$ probes per query in the limit.
},
even less than one. Probing  
consumes a small but nonzero amount of server CPU and 
network bandwidth, so it is desirable to lower the
probing rate to the extent that doing so has only negligible impact on tail latency. 
Our experiments in \Cref{sec:tunable} reveal
that $\rprobe > 1$ is
quite safe from the standpoint of tail latency. 

\paragraph{Load signals}
\resub{\plb includes a server-side module for tracking RIF and latency statistics and
responding to probes, as follows. 
We say that the query \emph{arrives} at the server when the application 
logic receives the RPC from Stubby,
and \emph{finishes} when the application logic hands the response RPC back to Stubby.
We define the latency of the query to be the length of this interval, during which 
the query contributes to this server's RIF count. If there is any application-level 
queueing, then the latency includes the sojourn time in the queue. However, it is 
more common for the application to eschew queueing and rely on thread or fiber scheduling instead.
We do not attempt to capture the network latency, because all server replicas reside in the same datacenter.

When responding to a probe, the RIF comes from simply checking a counter. The latency estimate is more subtle.
When a query finishes, we record its latency, tagged by the value of the RIF counter when it arrived. When a probe prompts us to estimate latency, we consult a set of recent latency values at (or near) the current RIF, and report the median. The per-query overhead of maintaining these data structures is small.
}
%
\aadelete{The per-query overhead of maintaining this state
on the \bkedit{replica} is small: updating all of the state 
variables requires only a small constant number of 
operations. 
When answering a probe, the responder first 
checks the current RIF value and then computes 
a latency estimate using the \bkedit{median} of 
recent latency measurements at the same or nearby RIF 
values. The latency estimation procedure is constrained 
to use samples from queries that arrived in the past 
second, but}
At moderate-to-high query arrival rates, 
the samples are plentiful enough that we base the latency 
estimates entirely on queries that finished
in the last O(1-10) milliseconds.

\paragraph{The probe pool}

\plb clients maintain a pool of 
probe responses to be used in replica selection. Each pool element indicates
the replica that responded, the response receipt
time,\footnote{The sent time would be ideal, but could introduce clock skew.} 
 and the load signals discussed above. The pool is capped at a maximum 
size: we have found that a pool size of 16 suffices 
to achieve the benefits of \plb, and the gains 
from increasing beyond 16 
are modest. New probe responses are added to the pool upon receipt, evicting a 
probe if necessary to respect the maximum pool size. 

\paragraph{Replica selection}

\plb could access many load signals, and we have chosen to focus on RIF and latency. 
The latency signal is obvious: since we want to minimize latency, why not route queries 
to the replicas exhibiting minimum latency? The RIF part is important because when the 
replica must store significant per-query state (as in YouTube), its RAM allocation must 
be a constant offset plus a term that is linear in its max anticipated RIF. In 
addition, RIF is an instantaneous signal that is a leading indicator of future load, 
since it represents the queries processing \emph{right now}. The latency signal is 
based on observed latencies of queries that have completed \emph{recently}. 
\bkedit{Both signals are more up-to-date than 
CPU utilization, which must be averaged over a long time period to be meaningful.}

From the RAM perspective, RIF matters only if our choice of replica raises the max RIF. 
In principle, we would prefer to balance on latency at all other times, although we 
must temper this impulse because of the leading versus trailing indicator 
consideration described above. 

To minimize both latency and RIF, \plb 
\resub{selects replicas using a 
{\em hot-cold lexicographic} (HCL) rule}
that labels probes in
the pool as either \emph{hot} or \emph{cold} based
on their RIF value. 
\plb clients maintain an estimate of the distribution 
of RIF across replicas, based on recent probe 
responses. They classify pool elements 
as hot if their RIF \bwdelete{equals or }exceeds a specified quantile (\rifpercent) 
of the estimated distribution, otherwise cold.
In replica selection, if all probes in the pool are hot, 
then the one with lowest RIF is chosen; otherwise, the cold probe with the lowest
latency is chosen. Our experiments (\Cref{sec:selection-rule}) suggest that 
$\rifpercent \in [0.6, 0.9]$ is a good choice, although even 0 is effective (\ie RIF-only control).

An exception arises if the pool is empty. 
Then, \plb simply falls back to selecting a uniformly random 
replica. In fact, our experience suggests it is useful to invoke this fallback whenever the pool occupancy drops below 2.

One popular approach to combining two signals is to 
use a linear combination of them with \bkedit{a} scaling constant to put them 
into the same units.
\aadelete{In our experience here and in many other contexts, such an 
approach is tough to apply well in practice because it can be difficult to 
select the right scaling constant and update it as circumstances change, whereas 
the hot-cold approach is more robust.}
\resub{In our experiments, \hotcold outperforms RIF-only (\Cref{sec:tunable}), which outperforms every non-trivial 
linear combination of RIF and latency 
(\Cref{sec:selection-rule} and 
\Cref{sec:lincombo}).}
The \hotcold approach also nicely captures 
our hierarchy of concerns: containing latency is nice, but replicas obeying their RAM 
allocations is often a hard constraint.


\paragraph{Probe reuse and removal}

\plb manages the probe pool to avoid three
\bkedit{conditions}: 
\emph{staleness}, when the 
load signals are too old to be accurate;
\emph{depletion}, when the pool becomes empty; and
\emph{degradation}, when the loads represented in the pool exhibit a selection bias towards replicas with higher load.
These are intertwined and the discussion will necessarily be somewhat 
circular, so we start with the simplest.

\emph{Depletion:} \plb employs several probe removal processes (detailed below) to 
control staleness and degradation, 
including removing probes when they are used. In order to stave off pool depletion without 
increasing the probing rate, we can extend the life of each probe by reusing it up to \reuselimit times. This reuse limit is set according to the formula
\begin{equation}
\label{eq:probe-reuse-limit}
    \reuselimit = \max \left\{ 1, \tfrac{1+\delta}{(1-m/n) \cdot \rprobe - \removeworst} \right\} ,
\end{equation}
where $\delta > 0$ is a configuration parameter that 
governs the net rate at which probes accumulate 
in the pool, $m$ is the maximum pool size, $n$ is 
the number of replicas, $\rprobe$ is the probing rate
discussed above, and $\removeworst$ is the rate of probe
removal discussed below. We always set $\reuselimit \geq 1$, 
and when it is fractional, we randomly round it to its floor or ceiling so as to preserve the expectation.

\emph{Staleness} occurs for two reasons: aging and overuse.
As a probe ages, its load data becomes less accurate because 
\bkedit{the replica receives new 
queries and finishes ones it already had}. Overuse is a special case of aging that we can 
partially mitigate; namely, \bkedit{when the client \emph{itself} sends a query to that replica,
it can compensate by incrementing the RIF value on that probe}. Ideally, we would 
also increase its latency estimate, but currently we do not. Part of \plb's solution for 
general staleness is to set a time
limit on \bkedit{probes and remove them} from the pool
when their age exceeds this limit. In addition, whenever a new probe arrives that would
increase the pool beyond its size limit, we 
drop the oldest probe.


\emph{Degradation} is the subtlest of these phenomena.
If the probes corresponding to lightly-loaded replicas 
are continually being selected and removed
from the pool, then the probes that
remain after many rounds of replica
selection correspond to highly-loaded replicas.
To avoid this, \plb periodically
removes the worst probe from the pool. This is 
analogous to --- but much more permissive than ---
the use of only the least-loaded of 
$d$ randomly selected bins in the standard 
power-of-$d$-choices model. It turns out
that broadening the bin-selection policy from 
``use only the best of $d$ random probes''
to ``avoid using the worst of $d$ random 
probes'' qualitatively preserves the
theoretical guarantees for the power-of-$d$-choices
model~\cite{park2011generalization}.
When removing the worst probe, \plb alternates 
between two rules: removing the oldest probe (\ie worst age) and 
removing the probe deemed worst according to the same ranking 
used for replica selection (but in reverse): if at least one probe
in the pool is hot, the hot probe with highest
RIF is removed; otherwise, the cold probe with
highest latency is removed.

We define a \removeworst parameter, and delete that many probes 
from the pool with each query. As with the probing rate, 
this number may be fractional, in which case the
actual number of probes removed is always either
the floor or the ceiling, deterministically rounded
to achieve the configured rate of probe removal on 
average. By alternating our removals between oldest and most loaded,
we achieve a unified approach to avoid both staleness
and degradation (on top of the probe timeouts). 

To summarize, \plb removes probes from the pool 
for four reasons.
\bkedit{ It evicts the oldest probe in the pool
     when necessary to avoid exceeding the maximum
     pool size.
     Probes are removed once they reach their
     reuse budget or when their age
     exceeds a timeout threshold. 
     Additionally, they are removed 
     at a 
     configurable rate per query, alternating between worst and oldest.}

\paragraph{Synchronous mode}
So far, we described \plb with asynchronous probing (\ie async mode), which maintains 
a probe pool to take probing off the critical serving path. The asynchrony is helpful 
in many use cases, such as when the probe latency is non-negligible or the CPU overhead 
of probing is high compared to the queries. However, we have also implemented a sync 
mode of \plb in which there is no probe pool. Instead, when a query arrives, the client 
issues \aadelete{a specified number of} \resub{some $d$} probes (\resub{at least 2, typically 3-5}\aadelete{two, typically three to five}) 
to \resub{random} replicas \aadelete{selected uniformly at random}, waits to receive a sufficient number of responses (typically \resub{$d-1$}
\aadelete{one fewer than the number of probes issued}) and then chooses among \resub{those} \aadelete{the responses received} 
using the same replica selection rule described above. We describe the sync mode here 
because it has been used in YouTube services as reported in \Cref{sec:youtube}, but all 
of the experiments reported in  \Cref{sec:evaluation} use async \plb because 
that mode is better for most uses.

One significant use case that requires sync mode is when replicas hold state that 
influences the cost of query execution, \eg a replica may cache certain data in memory to 
prevent reading from a slower storage layer. Sync probing allows us to include relevant 
information from the query in the probe. If the replica then determines that it can execute 
that query more efficiently because of data it already has in the cache, then it can 
manipulate its reported load so as to attract the query\resub{, \eg by scaling down its reported load by 10x}. 
We have used sync \plb in this way for part of YouTube.

\paragraph{Error aversion to avoid sinkholing}


Suppose a certain replica has a problem (such as misconfiguration) that causes 
it to process queries very quickly by immediately returning errors for a 
non-trivial fraction of its queries. Then its latency on the remaining 
successful queries, RIF, CPU utilization, and other metrics will make it appear less 
loaded than it normally would, given the amount of traffic sent its way. If the load 
balancer is not smart about this, this replica can attract more and more traffic, 
in a phenomenon known as \emph{sinkholing}. \plb includes some heuristics to avoid 
sinkholing, but since they are not central to our contribution, we have chosen to 
simplify our exposition by omitting details.



\section{Testbed evaluation}
\label{sec:evaluation}

\resub{In \Cref{sec:youtube}
we reported on our experience
deploying \plb in YouTube. The
comparison between \plb and \wrr in that section constitutes one type of evaluation, under conditions that arguably represent the ground-truth definition of a realistic workload. However, evaluating the deployed version of \plb in production does not shed light on how individual aspects of its design affect its performance. In this section we report on a set of tests
\bwcomment{microbenchmark in my mind is too small => set of controlled tests} 
performed on a testbed environment that allows for more controlled experiments with \plb and baselines, while preserving key features of the production environment for which \plb was designed: variable query costs and unpredictable time-varying antagonist load. 
}

\bkdelete{In this section, to illustrate each of \plb's 
interesting design features, we present the results of four experiments on a testbed, \emph{not} YouTube.}
We first describe our testbed and the baseline 
system parameters. Within each experiment, we explain which parameters we changed 
from the baseline.

Our testbed consists of one client job and one server job, each comprising 
multiple replicas running on distinct but identical physical machines colocated in the same 
datacenter. The queries represent a very simple CPU-intensive workload: they 
simply iterate an expensive hash function. In order to simulate 
variability in query costs, we vary the number of iterations, drawing it from a normal distribution 
whose standard deviation equals its mean (then truncated at zero). 
The jobs are running within a standard 
Google datacenter, on commodity multicore machines, using Google's standard 
isolation mechanisms, and the antagonist traffic is just whatever we happen 
to encounter in the wild.


All experiments use 100 client replicas and 100 server replicas, and each of the 
server replicas is allocated 10\% of the machine's CPU. In addition, probe pool size = 16, probes 
age out of the pool after one second, and the net probe-pool drift rate in \Cref{eq:probe-reuse-limit} is $\delta = 1$.
Unless otherwise specified, we set \rifpercent = $2^{-0.25} \approx 0.84$ and \removeworst = 1. We use 3 probes per query as our baseline probe rate to stay safely away from probe rates low enough to impact performance.
We target different aggregate  
utilizations in each experiment as necessary to induce the behaviors we wish 
to illustrate. Wherever we plot CPU utilization, it is scaled as a percentage 
of the allocation.

Some of the plots in these experiments show quantiles of RIF values. Since this data is all integer, 
the quantiles should be integer as well. However, when our monitoring system builds histograms, all 
instances of an integer $k$ are uniformly smeared across the interval $[k-\half,k+\half)$. For this 
reason, the plots of our RIF quantiles contain fractional values. \bkcomment{After deleting stuff that may need to go due to space constraints, it's possible the probing rate experiment will be the only plot showing RIF quantiles. If so, we should defer this discussion to that section.}

\subsection{Robustness to variable antagonist load}
\label{sec:wrr}


\begin{figure}[!t]
    \centering
    \begin{tikzpicture}[scale=0.98]
        \draw (-0.03,7.5) node {\includegraphics[width=0.48\textwidth]{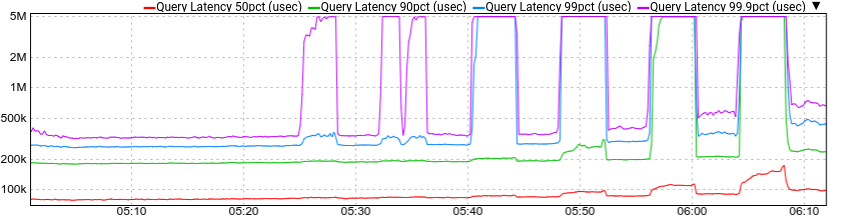}};
        \draw (0.07,4.25) node {\includegraphics[width=0.49\textwidth]{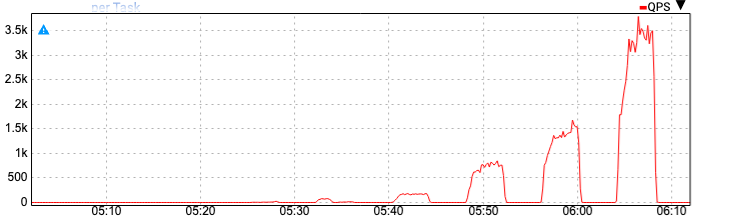}};
        \draw (-0.035,1) node
        {\includegraphics[width=0.485\textwidth]{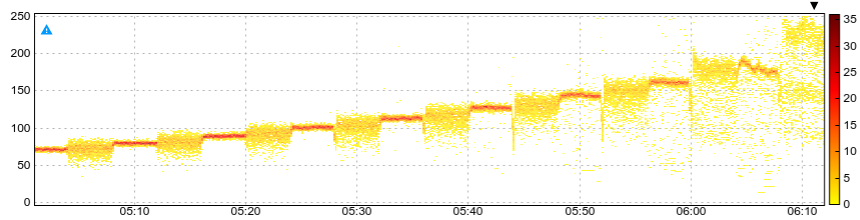}};
        \def\sw{0.455} 
        \def\xbase{0.36}
        \def\ramplvl{{75,83,93,103,114,127,141,157,174}}
        \foreach \i in {-5,-4,-3,-2,-1,0,1,2,3} {
        \fill[ultra nearly transparent]
        (\xbase + 2*\sw*\i,-0.25) rectangle (\xbase + \sw + 2*\sw*\i,9.5);
        \fill[ultra nearly transparent]
        (\xbase + \sw + 2*\sw*\i,8.75) rectangle (\xbase + 2*\sw + 2*\sw*\i,9.5);
        \ifnum \i > -5 
            \draw[line width=0.25,gray!85] (\xbase + 2*\sw*\i,0) -- (\xbase + 2*\sw*\i,8.48);
        \fi
        \tikzmath{\nodelabel = \ramplvl[\i + 5];}
        \node at (\xbase + \sw + 2*\sw*\i,8.9) {\footnotesize \nodelabel\%};
        }
        \draw[dashed,line width=2pt,gray!50] (\xbase - 4*\sw,9.5) -- (\xbase - 4*\sw ,-0.25);
        \draw (-2.8,9.3) node {\bf \small Below Alloc $\longleftarrow$};
        \draw (-0.1,9.3) node {\bf \small $\longrightarrow$ Above Alloc};
        \draw (0,6) node {(a) Tail Latency};
        \draw (0.03,2.75) node {(b) Errors};
        \draw (-0.04,-0.5) node {(c) Distribution of CPU Utilization};
    \end{tikzpicture}
    \caption{Load ramp experiment. Gray background denotes WRR policy, white denotes \plb. Note that tail latency is plotted on a log scale.
    }
    \label{fig:load-ramp}
\end{figure}
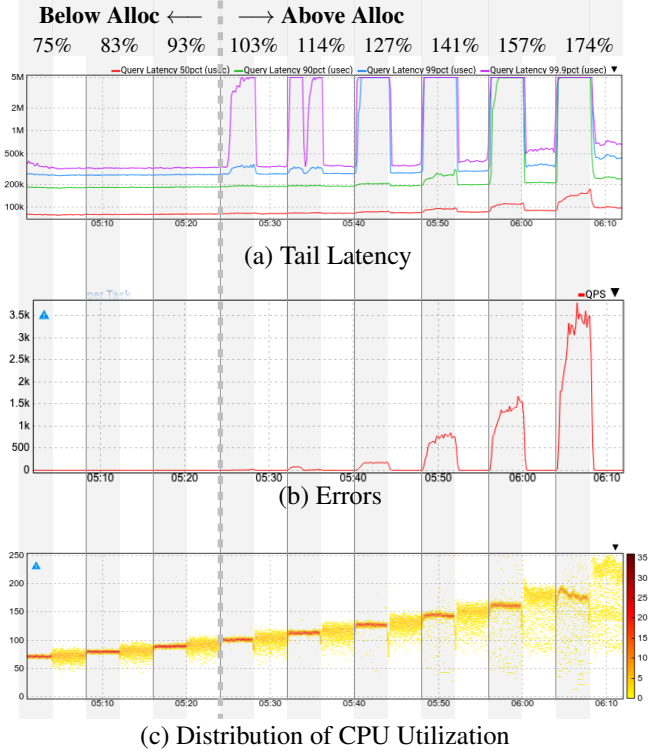


In this experiment, we start with the aggregate CPU load at about 75\% of 
our allocation, and ramp it up in 8 multiplicative steps of 10/9, 
yielding 0.75x, 0.83x, 0.93x, 1.03x, 1.14x, 1.27x, 1.41x, 1.57x, and 1.74x 
our allocation. \bkdelete{Thus, the first 3 load levels are below our allocation, and the 
last 6 are above.} We increased the load by increasing the aggregate query rate 
from about 5.6k queries per second (henceforth, qps) to around 13k qps, while holding the mean work per query 
constant. Within each load level, we use WRR for the first half of the period, 
then \plb for the second half.

The top plot in \Cref{fig:load-ramp} shows the effect on median and tail 
latency, on a log scale, measured in microseconds. We employ a 5s timeout for each query, 
so the graph tops out at 5s. In the first three steps (below allocation) the 
latency distribution stays steady for both WRR and \plb at roughly 80ms (p50), 
182ms (p90), 265ms (p99), and 325ms (p99.9). Then, as soon as we first exceed 
our CPU allocation (by 1.03x) in step 4, the behaviors of WRR and \plb rapidly diverge.

For WRR, the p99.9 latency maxes out at the 5s limit, while p99 jumps about 26\% 
to 335ms, slightly above the previous p99.9. By step 6 (1.27x allocation), p99 
has maxed out, and p90 is starting to show a visible bump to \roughly 203ms (+11\%). By step 8 (1.57x allocation), p90 has maxed out, and even p50 has risen 
to \roughly 111ms (+39\%). By step 9, even p50 has risen to \roughly 150ms (+88\%).

In contrast, \plb suffers almost no visible change at p99.9 in load steps 4 and 5, 
and by step 6, p99.9 rises only to \roughly 350ms (+8\%). It does begin to rise appreciably 
at load step 7 (1.41x allocation), but even at step 9 it is still contained around 
700ms, far from the 5s timeout. Meanwhile p99 latency does not visibly degrade 
until load step 7, p90 does so around step 8, and p50 does so around step 9.

The second plot shows error rates, which are zero for both WRR and \plb up through 
step 3. Note that this is the \emph{absolute} number of errors per second, not a 
percentage. WRR starts showing a very small error rate (5 to 20 errors/s) at load 
step 4. By load step 5, the error rate rises visibly, increasing much faster than 
the incoming qps. By step 9, more than a quarter of all queries are returning errors. 
Essentially all of these are "deadline exceeded" errors, from queries that hit their 5s deadline.
In contrast, \plb returns \emph{zero errors} at all load levels in this experiment.

The bottom plot shows a heatmap of the CPU utilization distributions for WRR and \plb. 
Here, one can see that WRR is doing a superb job at accomplishing what it was designed 
to do: balance CPU load. In contrast, the CPU distribution for \plb is substantially 
looser. So why are the latency and error metrics for WRR so bad? It is precisely the explanation that we offered as motivation in \Cref{sec:motivation}. Digging 
further into the server-level data for WRR (not pictured), we see a strong correlation between the 
server replicas with high antagonist load and ones where our latency suffers. Meanwhile, \plb 
shifts load away from those server replicas to avoid the worst of the latency impact.

This collection of results explains the title of our paper. Although it is 
counterintuitive for most people, in this case the load balancer that achieves 
near-perfect load balance (WRR) is clearly worse (higher errors and latency). 
That is because \emph{the real goal of a load balancer is not to balance load,} \bkedit{it is to direct load where capacity is available.} 
\srcomment{Didn't we show earlier that WRR is far from near-perfect (1s vs. 1m plots)?}

\bkdelete{Does it mean we can pack our services 
so that all machines are 100\% allocated, and then \plb will allow us to run each of 
these services at 1.41x their allocations with no ill effects? No, because \plb does 
not manufacture CPU cycles out of thin air.}
These results
are possible only because 
many server replicas in our system are \emph{not} fully allocated with antagonists, and 
at any given moment, many of these antagonists are not using their full CPU allocations. 
\plb allows our job to shift load among replicas so as to fit into these cracks of 
temporary spare capacity. This has two implications for service provisioning: 
(1) We can provision much more aggressively, because \plb can deal with transient 
load spikes. (2) In theory, this could even enable overcommitment, because 
it enlarges the relevant resource pool from the level of one machine to a much larger 
set of machines, enabling more effective statistical multiplexing.

\subsection{Replica selection rule}
\label{sec:selection-rule}

In modern datacenters with thousands of machines, it is common for the machines to span multiple hardware generations and processor architectures. When provisioning a job across multiple 
machines, we can try to account for these differences by applying a performance 
multiplier to different machine types, \eg 1 CPU core of type A is equivalent to 1.2 
CPU cores of type B. However, it gets tricky to do this well at scale because 
the performance skew is often highly dependent on the particular workload, \eg 
YouTube may be more suited to one of the hardware generations, while Google Maps is better suited to the other. 
As a result, when our replicas are assigned to machines with different hardware, 
their effective throughput becomes heterogeneous, and it is 
the faster machines that exhibit lower 
latency when unloaded (by definition). In this case, the way to minimize mean latency is to fill the fast machines 
until their latency degrades to match the slow machines, and then fill both fast and slow together.

\resub{
    There are various ways that a replica selection rule
    could be designed to fulfill this goal. We experimented
    with nine replica selection rules ranging from very
    simple baselines to sophisticated rules used in 
    popular open-source load balancers (e.g.~\cite{YARP-homepage})
    and in the research literature (e.g.~\cite{suresh15c3}).

    In describing the replica selection rules we evaluated, 
    we distinguish between {\em client-local} and 
    {\em server-local} quantities. The former are 
    measured at the client (or load balancer) itself, 
    the latter are measured at the server replica 
    and reported back to the client (or load balancer). 
    For example, client-local RIF refers to the number
    of queries the client has sent to the replica that
    have not yet yielded responses. Server-local RIF refers to 
    the total number of queries the replica has received 
    (from all clients) that it has not yet finished processing.
    \bkcomment{Someone please check that the following
    discussion of the connection vs.\ request distinction
    is OK.}
    In the literature on load balancing of HTTP traffic
    there is also a distinction between individual requests
    and connections; the latter may be long-lived, comprising 
    multiple requests. Since the queries in our work represent short-lived connections, we ignore
    this distinction in our experiment, treating connections
    and requests as synonymous.
    
    We evaluated the following replica selection rules.
    \begin{itemize}[itemsep=0pt]
        \item {\bf Random} selects a uniformly random replica.
        \item {\bf Round Robin (RR)} cycles through the replicas,
            keeping track of the most recently chosen one and always selecting the next available replica in cyclic order.
        \item {\bf Weighted Round Robin (WRR)} is as described in \Cref{sec:motivation}.
        \aadelete{is similar to RR, 
            but it is designed to account for (transient) differences in processing speeds. It maintains an estimate of cost per query on each replica (updated using real-time measurements) and modulates the frequency of choosing replicas to equalize their estimated cost rather than the number of queries they process.} 
        \item {\bf Least Loaded (LL)} represents the LeastLoaded policy implemented in 
            the NGINX and Envoy reverse proxies~\cite{nginx-load-balancing,envoy-github}. 
            It chooses the available 
            replica with the least client-local RIF, breaking ties in favor of one nearest to the most-recently-chosen replica in cyclic order.
        \item {\bf Least Loaded with Power of Two Choices (LL-Po2C)} 
            samples two available replicas uniformly at random and 
            selects the one 
            with the least client-local RIF. This modification
            of LL is also implemented in NGINX and Envoy. 
        \item {\bf YARP-Po2C} is a replica selection rule from Microsoft's YARP 
            reverse proxy library~\cite{YARP-homepage} based on power-of-two-choices. 
            All replicas are periodically 
            polled to report their (server-local) RIF. Replica selection is 
            performed by randomly sampling two replicas and
            selecting the one with lower reported RIF. In our
            experiments we set the polling interval to 500ms, 
            a 30x faster rate of polling than in the standard YARP-Po2C
            implementation. We chose the 500ms interval to approximately equalize the total number of RIF 
            reports each client receives, per second, 
            with the number of probe responses received
            per second by \plb clients in
            our experiment.
        \item {\bf Linear, C3, and \plb} all use the asynchronous
            probing method described in \Cref{sec:design}, but 
            they differ in the scoring rule used to select a 
            replica from the pool of probe responses.
        \item {\bf Linear} uses a linear combination of 
            RIF and latency. To represent RIF and latency in
            comparable units, we scale RIF by the median 
            query processing time measured on replicas with
            one request in flight. A replica's score is defined
            to be an equally weighted average\footnote{In the supplementary material, we 
            report on an experiment showing that the performance
            of the Linear rule is equally poor for most weightings, except when
            it degenerates to RIF-only control.} of latency and
            scaled RIF.
        \item {\bf C3} in this paper uses the replica scoring function described
            in~\cite{suresh15c3} with \plb's probing logic. It computes a RIF estimate for each replica as $\hat{q} = 1 + os \cdot n + \bar{q}$, where $os$ is the client-local RIF, $n$ is the number of clients participating in the job, and $\bar{q}$ is an exponentially weighted moving average of the server-local RIF. It then computes a score for each replica as $\Psi = (R - \mu^{-1}) + \hat{q}^3 \cdot \mu^{-1}$, where $R$ and $\mu^{-1}$ are exponentially weighted moving averages of the client-local and server-local response time, respectively.
        \item {\bf \plb} uses the \hotcold replica selection 
            rule described in \Cref{sec:design}, with the 
            RIF limit quantile set to $\rifpercent = 0.75$. 
    \end{itemize}
\definecolor{bar0}{rgb}{0.6,0,0}
\definecolor{bar1}{rgb}{0.6,0.3,0}
\definecolor{bar2}{rgb}{0.8,0.6,0}
\definecolor{bar3}{rgb}{0,0.6,0}
\definecolor{bar4}{rgb}{0,0.4,0.8}
\definecolor{bar5}{rgb}{0.5,0,0.5}
\definecolor{bar6}{rgb}{0.4,0.4,0.6}
\definecolor{bar7}{rgb}{0,0.5,0.5}
\definecolor{bar8}{rgb}{0.5,0.5,0.5}

\newcommand{\rulename}[1]{
    \ifnum #1 = 0
        RoundRobin
    \fi
    \ifnum #1 = 1
        Random
    \fi
    \ifnum #1 = 2
        WeightedRR
    \fi
    \ifnum #1 = 3
        LeastLoaded
    \fi
    \ifnum #1 = 4
        LL-Po2C
    \fi
    \ifnum #1 = 5
        YARP-Po2C
    \fi
    \ifnum #1 = 6
        Linear
    \fi
    \ifnum #1 = 7
        C3
    \fi
    \ifnum #1 = 8
        \plb
    \fi
}

\def\lspn{{4984,294,173,343,224,210,206,161,149}}
\def\lspnn{{5026,5023,314,1804,569,2642,698,299,281}}
\def\lnpn{{5013,5007,1667,940,623,213,245,164,152}}
\def\lnpnn{{5029,5028,5001,2654,1918,1169,1036,304,286}}

\newcommand{\vertellipsis}[1]{\foreach \yve in {12.8,13.1,13.4} { \filldraw[black] (#1,\yve) circle (0.05); } }
\newcommand{\horizellipsis}[1]{\foreach \xve in {8.9, 9.1, 9.3} {
\filldraw[black] (\xve,#1) circle (0.02); } }

\begin{figure}[htb]
    \centering
    \begin{tikzpicture}[scale=0.75]
        \foreach \i in {2,4,6,8,10,12}
        {
            \tikzmath{\x = 0.7*\i;}
            \draw[dashed, line width=0.5] (\x,11) -- (\x,0.5);
            \draw (\x,11.1) -- (\x,10.9);
            \draw (\x,11.4) node {$\i00$};
        };
        \foreach \i in {0,1,2,3,4,5,6,7,8} {
            \tikzmath{
                \xf = 0.007; 
                \barst = 10.5 - 0.5*\i;
                \barsb = 10.0 - 0.5*\i;
                \smw = {min(\lspn[\i],1250)};
                \srw = {min(\lspnn[\i],1250)};
                \barsm = \xf * \smw;
                \barsr = \xf * \srw;
                \barnt = 5.5 - 0.5*\i;
                \barnb = 5.0 - 0.5*\i;
                \nmw = {min(\lnpn[\i],1250)};
                \nrw = {min(\lnpnn[\i],1250)};
                \barnm = \xf * \nmw;
                \barnr = \xf * \nrw;
                \smval = \lspn[\i];
                \snval = \lspnn[\i];
                \nmval = \lnpn[\i];
                \nnval = \lnpnn[\i];
            }
            \fill [bar\i] (0,\barst) rectangle (\barsm,\barsb);
            \fill [bar\i!25] (\barsm,\barst) rectangle (\barsr,\barsb);
            \fill [bar\i] (0,\barnt) rectangle (\barnm,\barnb);
            \fill [bar\i!25] (\barnm,\barnt) rectangle (\barnr,\barnb);
            \node at (\barsm - 0.55,10.25 - 0.5*\i) {\small \bf \textcolor{white}{\textsf{
                \ifnum \smval < 5000 \smval
                \else TO \fi}}};
            \ifnum \snval < 5000 
                \node at (\barsr - 0.4, 10.25 - 0.5*\i) {\small \bf
                \textsf{\snval}}
            \else
                \node at (\barsr + 0.3, 10.25 - 0.5*\i) {\small \bf
                \textsf{TO}}
            \fi;
            \node at (\barnm - 0.55,5.25 - 0.5*\i) {\small \bf \textcolor{white}{\textsf{
                \ifnum \nmval < 5000 \nmval
                \else TO \fi}}};
            \ifnum \nnval < 5000 
                \node at (\barnr - 0.4, 5.25 - 0.5*\i) {\small \bf
                \textsf{\nnval}}
            \else
                \node at (\barnr + 0.3, 5.25 - 0.5*\i) {\small \bf
                \textsf{TO}}
            \fi;
        };
        \draw[->] (0,11) -- (0,0.5);
        \draw[->] (0,11) -- (9,11);
        \draw (4.2,11.9) node {\bf  Latency (ms)};
        \draw (-0.5,8.5) node {\bf  70\%};
        \draw (-0.5,3.5) node {\bf  90\%};
        \draw (-1,6) node[rotate=90] {\bf Load (\% of alloc)};
        \foreach \i in {0,1,2,3,4,5,6,7,8} {
            \tikzmath{ \legbx = 6.8; \legby = 3.2 - 0.5 * \i; }
            \ifnum \i < 6 
                \tikzmath{ \legbx = 3.3; \legby = 1.7 - 0.5*\i; }
            \fi
            \ifnum \i < 3
                \tikzmath{ \legbx = -0.2; \legby = 0.2 - 0.5*\i; }
            \fi
            \fill [bar\i] (\legbx, \legby) rectangle (\legbx + 0.7, \legby - 0.2);
            \node at (\legbx + 2.35, \legby-0.1) {\parbox{2cm}{\small \rulename{\i}}};
        };
    \end{tikzpicture}
    \caption{Comparison of replica selection rules. Dark portions of bars represent 90th percentile, light portions are 99th percentile. Bars representing latencies exceeding 1.25sec are truncated. Timeouts are represented by 'TO'.}
    \label{fig:bar-graph}
\end{figure}
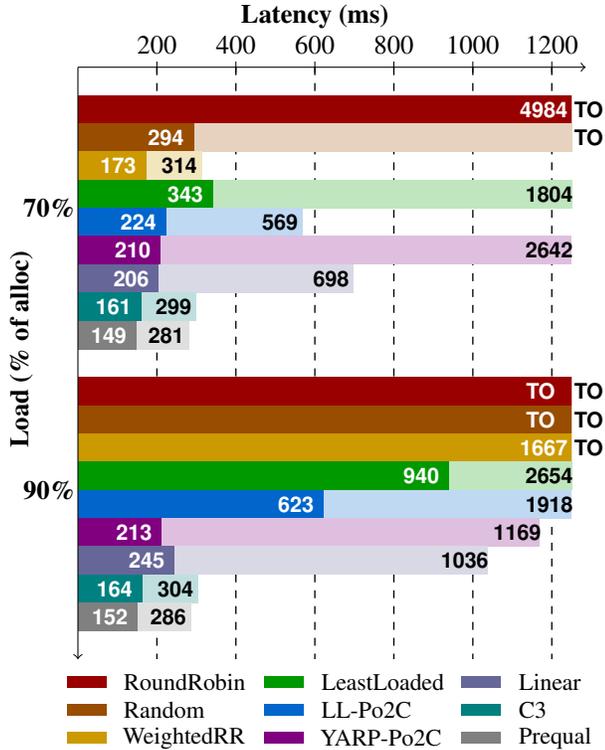
The results of our experiment are depicted 
in \Cref{fig:bar-graph}. We tested the replica 
selection rules at two different levels of 
aggregate CPU
load --- 70\% and 90\% of our allocation --- 
and reported two quantiles of tail latency, 
the 90th percentile (p90) and 99th percentile (p99). 

The replica selection rules that performed best
at all load levels and latency quantiles were 
C3 and \plb. What these rules have in common is
that they incorporate server-local 
quantities, they penalize high (server-local) RIF severely, 
and they favor low-latency replicas when there are 
multiple replicas with low RIF. In the case of
\plb this behavior is evident in the way it 
distinguishes between hot and cold replicas.
In the case of C3 it is implicit in the scoring
function's cubic dependence on estimated queue
size, $\hat{q}$: when $\hat{q}$ is near zero it
contributes negligibly to the score, but it 
rapidly grows to predominate the score as 
$\hat{q}$ moves away from zero. 

Comparing C3 and \plb, one sees a small 
quantitative advantage for \plb (3-8\%) 
at all load levels and latency quantiles 
we tested. Another convenient feature of 
\plb's replica selection rule is that it 
has a tunable parameter, \rifpercent,
that allows it to span the full range from
RIF-only to latency-only control, depending
on the needs of the application. Additionally,
\plb uses fewer signals than C3, simplifying
the implementation and monitoring.

It is noteworthy that the LL policy, 
which bases replica selection on client-local 
rather than server-local RIF, 
experiences high p99 latency 
when load is at 70\% of allocated capacity.
Even when a server has no active connections 
from a given client, it can be 
highly loaded with queries from other clients.
If this possibility happens more than 1\% of 
the time, that is enough to impact p99 latency. 
When load is 90\% of allocation, we see that
even p90 latency suffers heavily under
the LL policy. Combining LL with Po2C 
improves latency at all load levels and
quantiles in our experiments, but the LL-Po2C 
rule still lags far behind \plb and C3.

YARP-Po2C's selection rule using server-local 
RIF fares better at high load than LL-Po2C
which uses client-local RIF, as one might expect. 
However, its decisions are often based on stale 
information due to the infrequent polling of 
replicas, and this adversely affects latency.

Interestingly, the replica selection rule based on a 50-50 linear combination of latency and RIF performed much worse than \plb's and C3's scoring rules. 
This indicates that a linear 
function of RIF doesn't penalize high RIF severely enough compared to C3's cubic function or \plb's strict prioritization of cold replicas over hot ones.

Finally, the WRR policy performed quite well when load was at 70\% of allocation, but its p99 latency suffers greatly at 90\% load. This is consistent with the results reported in \Cref{sec:wrr}, where WRR experiences a very sharp increase in tail latency in response to a modest increase in aggregate CPU load. (WRR experienced this crossover at different load levels in the two experiments, probably because of differing amounts of antagonist load.) \bkcomment{A skeptical reader might observe that the variability of antagonist load calls all of our experimental results into question: if configuration A had better tail latency than configuration B, who's to say that's not because the antagonist load was greater during the time B was running? An appropriate response is that for each of the experiments we're reporting, we cycled through the configs multiple times, and the comparisons we're reporting were qualitatively consistent across cycles of the experiment, even if there were quantitative differences. Do you think it's appropriate to address this issue, e.g. in the opening portion of \Cref{sec:evaluation} before we get to \Cref{sec:wrr}? None of the reviews raised this criticism of our experimental design, so I'm leaning toward ignoring the issue.}
}

\subsection{Tunable parameters}
\label{sec:tunable}

\resub{
\plb has a few tunable parameters that influence its 
behavior and performance. In this section we evaluate
the impact of two of those
parameters, probe rate and RIF limit threshold. The
first influences the freshness of the load signals
used for replica selection, while the second 
influences the selection rule's propensity for
steering queries toward low estimated latency
versus avoiding high RIF. 
}

\begin{figure}[!t]
    \centering
    \begin{tikzpicture}[scale=0.9]
        \def\heightsquash{0.16\textheight}
        \draw (0,11.1) node {\includegraphics[width=0.48\textwidth,height=\heightsquash]{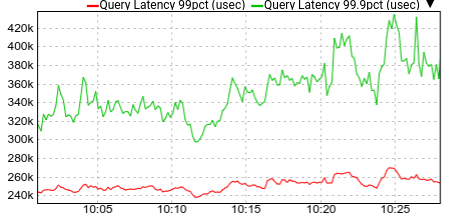}};
        \draw (0,8.7) node {(a) Tail Latency at 99p and 99.9p};
        \draw (-0.01,6.3) node 
        {\includegraphics[width=0.48\textwidth,height=\heightsquash]{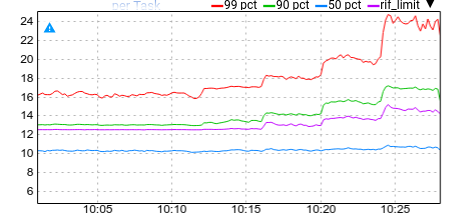}};
        \draw (0,3.9) node {(b) RIF Quantiles}; 
        \draw (-3.0,13.7) node {4};
        \draw (-2.12,13.7) node {$2 \sqrt{2}$};
        \draw (-1.0,13.7) node {2};
        \draw (0.09,13.7) node {$\sqrt{2}$};
        \draw (1.22,13.7) node {1};
        \draw (2.35,13.7) node {$\sqrt{\frac12}$};
        \draw (3.48,13.7) node {$\frac12$};
        \def\xbase{-2.68}
        \def\sw{1.12}
        \foreach \i in {0,2,4} {
            \fill[ultra nearly transparent]
            (\sw*\i + \xbase,4.3) 
            rectangle
            (\sw*\i + \xbase + \sw,14.5);
        }
        \foreach \i in {-1,1,3,5} {
            \fill[ultra nearly transparent]
            (\sw*\i + \xbase,13.3) 
            rectangle
            (\sw*\i + \xbase + \sw,14.5);
        }
        \draw (0,14.2) node {\bf Probe Rate};
    \end{tikzpicture}
    \caption{Probing rate experiment. Rates (top) are 
    expressed in probes/query.}
    \label{fig:probing-rate}
\end{figure}
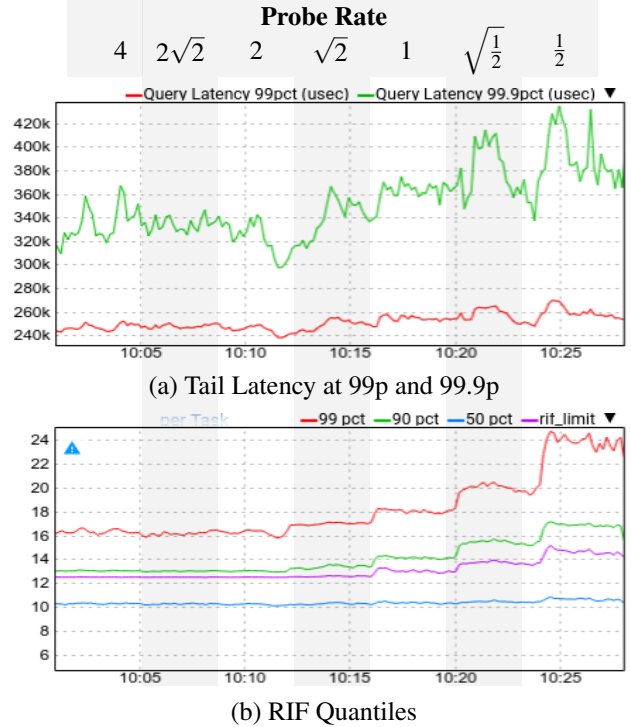


\paragraph{Probing Rate}
In the YouTube application, the probes are so light compared to the queries 
themselves that the probing overhead is negligible. Unfortunately, this \bkedit{is not} true 
for all applications. Thus, we seek to identify the minimum probing rate 
we can tolerate while retaining the benefits of \plb.

In this experiment, we ramp down the probing rate from 4x to \half x the query rate, 
in 6 multiplicative steps of $\sqrt{2}$ each, while keeping the probe removal rate 
steady at 0.25 per query, and \reuselimit increases to compensate, guided by \eqref{eq:probe-reuse-limit}.
In order to magnify the effects, we ran the system very hot, at roughly 
1.5x our CPU allocation throughout. 

\Cref{fig:probing-rate} shows the resulting latency and RIF, \bkdelete{(median and tail),} as well as the \riflimit control parameter. \bkdelete{Note that 
the p50 latency plot is zoomed in so much that the variation looks large even though 
the min and max differ by only 5\%.} The take-home point is that \plb is fairly 
insensitive to the probing rate until we drop below one probe per query, at which point the negative 
effects become significant. At probing rates of $\frac{1}{\sqrt{2}}$x and $\half$x, 
the tail RIF distributions jump visibly, and this change is echoed by \bkedit{both}
latency quantiles.
Anecdotally, we have observed this phenomenon across many similar experiments, 
always around 1 probe per query.


\paragraph{RIF Quantile}
We now present an experiment to
explore the RIF vs 
latency tradeoff expressed by the \bkedit{quantile that distinguishes hot from cold replicas in the \hotcold rule}. To do this, we designate all of the odd-numbered server replicas (1,3,...,99) as fast, 
and the even-numbered ones (0,2,...,98) as slow. However, we do not actually seek 
out servers from different hardware generations. Instead, when generating the queries, 
we artificially inflate the query work by a factor of 2 on the even replicas, 
thereby causing it to burn 2x the CPU cycles and act as if it were 2x slower. In addition, we vary 
the \rifpercent parameter over the course of the experiment, starting 
at 0 (meaning RIF-only control), then ramping it up from $0.9^{10} \approx 0.35$ to 0.9, 
in steps of $\frac{10}{9}$, then to 0.99, 0.999, and 1 (meaning latency-only control). 
Thus, there are 14 steps overall. Throughout this experiment, the mean load is held steady at about 75\% of the CPU allocation.
\bwcomment{latex artefact shows \rifpercent parameter as merged, also present in other places.}

Note that there is actually a discontinuity in the behavior of \plb between the last two steps. At \rifpercent = 0.999, the RIF limit is effectively the maximum RIF, which means that any replica tied for the max is considered hot. In contrast, when \rifpercent = 1, the RIF limit is $\infty$, and every replica is considered cold.

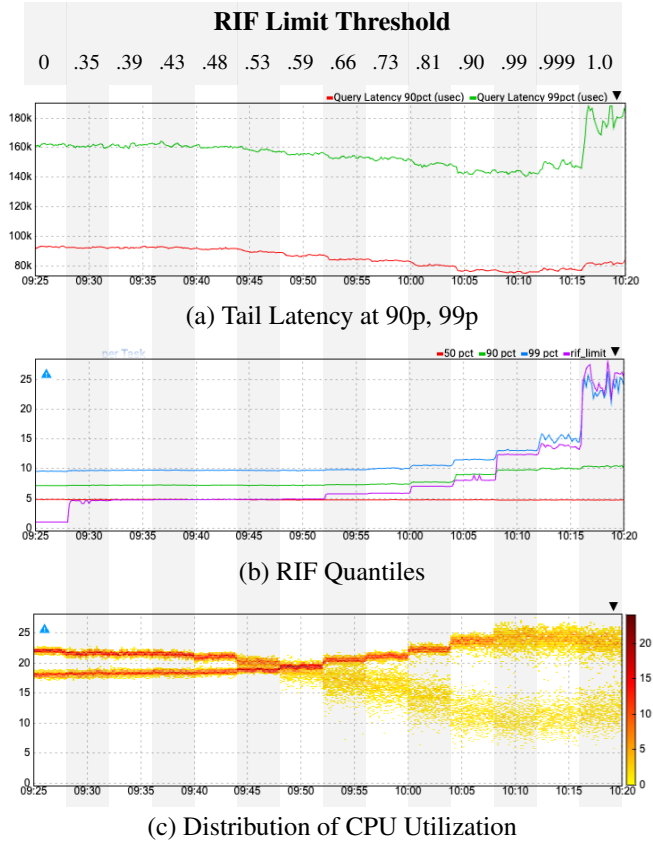
\begin{figure}[!t]
    \centering
    \begin{tikzpicture}
        \draw (-0.02,10) node {\includegraphics[width=0.482\textwidth]{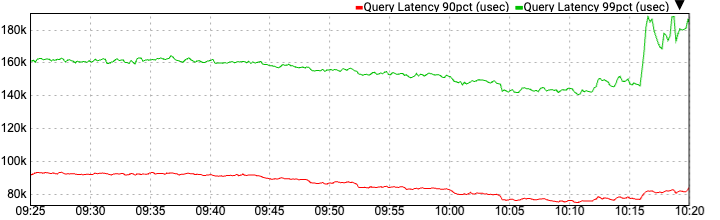}};
        \draw (0,8.3) node {(a) Tail Latency at 90p, 99p};
        \draw (0,6.6) node 
        {\includegraphics[width=0.48\textwidth]{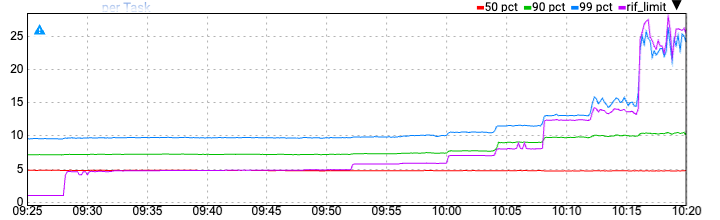}};
        \draw (0,4.9) node {(b) RIF Quantiles}; 
        \draw (0,3.2) node 
        {\includegraphics[width=0.48\textwidth]{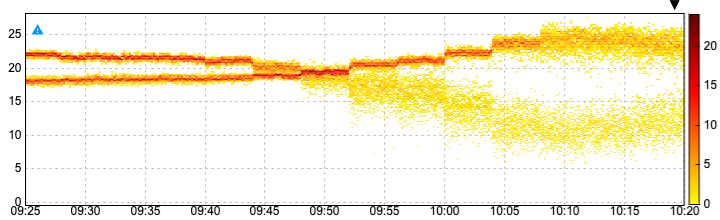}};
        \draw (0,1.5) node {(c) Distribution of CPU Utilization}; 
        \def\sw{0.568} 
        \foreach \i in {0,1,2,3,4,5,6} {
            \fill[ultra nearly transparent] (2*\sw*\i - \sw - 3.53, 11.4) rectangle (2*\sw*\i - 3.53, 12.5);
            \fill[ultra nearly transparent] (2*\sw*\i - 3.53, 1.8) rectangle (2*\sw*\i + \sw - 3.53, 12.5); 
        }
        \def\xoffs{-3.83}
        \def\yoffs{11.7}
        \draw (\xoffs + 0*\sw, \yoffs) node {\footnotesize 0};
        \draw (\xoffs + 1*\sw, \yoffs) node {\footnotesize .35};
        \draw (\xoffs + 2*\sw, \yoffs) node {\footnotesize .39};
        \draw (\xoffs + 3*\sw, \yoffs) node {\footnotesize .43};
        \draw (\xoffs + 4*\sw, \yoffs) node {\footnotesize .48};
        \draw (\xoffs + 5*\sw, \yoffs) node {\footnotesize .53};
        \draw (\xoffs + 6*\sw, \yoffs) node {\footnotesize .59};
        \draw (\xoffs + 7*\sw, \yoffs) node {\footnotesize .66};
        \draw (\xoffs + 8*\sw, \yoffs) node {\footnotesize .73};
        \draw (\xoffs + 9*\sw, \yoffs) node {\footnotesize .81};
        \draw (\xoffs + 10*\sw, \yoffs) node {\footnotesize .90};
        \draw (\xoffs + 11*\sw, \yoffs) node {\footnotesize .99};
        \draw (\xoffs + 12*\sw, \yoffs) node {\footnotesize .999};
        \draw (\xoffs + 13*\sw, \yoffs) node {\footnotesize 1.0};
        \draw (0,12.25) node {\bf RIF Limit Threshold};
    \end{tikzpicture}
    \caption{RIF Limit experiment. \rifpercent varies from 0 at left (pure RIF control) to 1 at right (pure latency control).}
    \label{fig:rif-limit}
\end{figure}
\Cref{fig:rif-limit} shows the results. The RIF limit threshold \rifpercent increases from 0 on the left (pure RIF control) to 1 on the right (pure latency control). The top two plots show the p99, p90, and p50 latencies. As we would expect, all three latency quantiles go down as we shift more towards latency-based control, up through step 12 (\rifpercent = 0.99). Specifically, p99 drops from \roughly 162ms to \roughly 142ms (-12\%), p90 drops from 93ms to 75ms (-19\%) and p50 drops from \roughly 34.5ms to \roughly 31ms (-10\%). At step 13, all three quantiles begin to edge up slightly. When we switch to full latency control in step 14, all quantiles move up sharply, especially p99, which jumps up from \roughly 148ms (step 13) to about 178ms (+20\%). Even more dramatic is p99.9 (not shown), which falls from 210ms in step 0 to 198ms in step 12, rises back to 210ms in step 13, and then fluctuates chaotically in the 337ms to 502ms range in step 14 (1.6x to 2.4x, compared to step 13).

\bkdelete{%
The third plot shows the p50, p90, and p99 RIF values, along with the RIF limit threshold, \riflimit, that governs the classification of replicas as hot or cold. The RIF quantiles are aggregated from direct observations by the server replicas themselves, whereas \riflimit is computed and reported by the clients. The salient observations here are that the p99 RIF is stable through step 9, then gradually rises in steps 10..12 but is flat within each of those epochs. In step 13, it rises modestly but is also somewhat unstable within that epoch. In step 14, it jumps dramatically from \roughly 15 to the 21-26 range.
}

It appears that even a tiny bit of RIF control goes a long way. Why is it that pure latency control results in such worse latency than  \rifpercent = 0.999? We hinted at our interpretation earlier: RIF is a valuable leading signal of load, so ignoring it entirely is a bad idea.

The bottom plot shows the CPU utilization distribution. 
Notice the two crossing bands, which correspond to the 
``slow'' (\ie even) and ``fast'' (\ie odd) replicas. The 
slow (resp. fast) replicas correspond to the band that 
is decreasing (resp. increasing) to the right.
This is exactly as we would expect, as increasing \rifpercent 
means we balance more often based on latency, which 
favors the fast replicas. In addition, the CPU distributions are tighter to the left, where RIF control is higher. This is because RIF is quite a good predictor of future CPU utilization. \bkdelete{We verified that if you scale down the slow band by a factor of 2, you get a plot that mirrors the qps sent to the fast and slow replicas. This matches the fact that the slow replicas have 2x the cost per query of the fast ones.}

This experiment justifies the \bkedit{\hotcold rule}, since turning the dial towards more latency-based control does indeed decrease latency, and turning it up even as high as 0.73 (step 9) has next to no impact on RIF. At step 7 (\rifpercent = 0.59), all three RIF quantiles are just as good as with RIF-only control, despite the fact that the vast majority of queries are being routed based on latency. In this experiment, the probe pool has size 16 and is nearly always full. Because of the use-best and remove-worst processes, the probe pool is not uniformly random. But if it were, then the probability of all 16 probes being hot would be roughly $2^{-16} \approx 1.5 \times 10^{-5}$, since \riflimit and p50 RIF are both 5 during step 7. Otherwise, the query is routed to a cold replica based on latency. This is what allows us to have the best of both worlds, most of the time: simultaneously routing based on latency \emph{and} avoiding the highest-RIF replicas. \bkdelete{\aaedit{This shows that the theoretical considerations sketched in the last paragraph of Sec 4.1 are substantiated in practice.}}




\section{Related work}
\label{sec:related}

The ``power of two choices'' paradigm for randomized load balancing was first analyzed in the highly influential work of Broder et al.~\cite{azar99balanced} and Mitzenmacher~\cite{Mitzenmacher96-phdthesis}. These ideas originated in theoretical work, but over the years they found their way into a wide variety of 
applications in industry, as noted in the prize citation for the 2020 ACM Paris Kanellakis Theory and Practice 
Award~\cite{Kanellakis-Award-2020-power-of-two-choices}. The original theoretical results on the power of two choices pertained to the 
case of throwing $n$ balls into $n$ bins, but they were soon generalized to the ``heavily loaded'' setting, when balls exceed 
bins~\cite{berenbrink06heavily}. 
Qualitatively, these results show 
that load-balancing processes 
based on randomly sampling multiple choices and selecting the best ones vastly outperform simpler procedures based on sampling a single 
bin for each ball. 
Subsequent theoretical research affirmed that this finding is surprisingly robust to varying the modeling assumptions or the 
specific rule for selecting among the sampled bins; see, for example, the surveys~\cite{mitz-survey,wieder-survey}. Of particular 
relevance to the design of \plb is the generalization to load-balancing processes with 
memory~\cite{mitzenmacher02memory,los22balanced,los23memory}, 
\resub{the closest analogue in the theory literature to our 
use of asynchronous probing.}

The need for datacenter task scheduling with sub-second latencies has prompted a great deal of 
research into decentralized load balancers. Several proposed systems in this branch of the load 
balancing literature leverage the power-of-$d$-choices \bkedit{(\podc)} paradigm. One prominent example is 
Sparrow~\cite{ousterhout2013sparrow}, a scheduler for highly parallel jobs that each consist of a 
large number of tasks. Sparrow aims to minimize the job response time, i.e.~the time when the last 
constituent task of a job finishes executing. This goal motivates a different set of design choices, 
combining batch sampling and late binding: a client whose job comprises $m$ tasks places $d \cdot m$ 
reservations on randomly selected server replicas; the $m$ replicas that become available soonest 
request to run the $m$ tasks, and the other $(d-1) \cdot m$ reservations are canceled. This achieves a 
similar effect to probing $d \cdot m$ replicas and selecting the $m$ best probe responses, with 
reservations playing the role of probes. Our setting differs in that jobs are not batched, which 
allows us to substitute a simpler and more efficient probing mechanism, avoiding the use of 
reservations and late binding which would incur undesirable overheads in our environment. Building 
upon Sparrow's batch sampling idea, Eagle~\cite{delgado2016eagle} is a hybrid scheduler that 
incorporates a distributed probe-based scheduler using batch sampling to handle short-lived tasks, 
alongside a centralized scheduler that places long tasks on replicas. The 
hybrid scheduler design makes 
sense in settings where a wide disparity in job sizes makes head-of-line blocking a potentially 
serious problem; in our setting the jobs (queries) are short-lived, allowing the fully distributed 
load-balancing solution offered by \plb to perform well in the production environment of YouTube 
and in our testbed.

A centralized \bkedit{\podc-based} load-balancer is incorporated into 
RackSched~\cite{zhu20racksched}, where it runs on a top-of-rack switch to make microsecond-scale inter-server 
scheduling decisions enabling rack-scale computing. Client-based probing solutions such as ours are 
inapplicable in their setting, where the resource cost of probing is comparable to the resource cost 
of serving requests. Instead of probing, their servers piggyback load signals into normal traffic and 
the scheduler treats these load signals as implicit probes.
\resub{%
Another rack-scale load balancing system using a 
RIF-based queueing policy is R2P2~\cite{kogias2019r2p2},
whose $\mathsf{JBSQ}(n)$ policy is a variation on the 
LeastLoaded policy discussed in \Cref{sec:selection-rule}.
Both RackSched and R2P2 rely on all connections passing
through a single router or switch; as explained in 
\Cref{sec:motivation}, at the scale of a service such as 
YouTube using a single load balancer is not a viable option.}

C3~\cite{suresh15c3} is another distributed load balancer with the same design goal as \plb --- minimizing tail 
latency by adaptively selecting replicas in response to real-time load feedback --- but a very 
different methodology using much more state. C3 uses distributed rate control and backpressure, 
with every client maintaining exponentially weighted moving average estimates of latency and RIF for 
every server replica, and 
\aadelete{limiting their sending rates} using a token-bucket based rate limiter for 
each server replica. 
\aadelete{The authors of~\cite{zhu20racksched} evaluated C3 on a deployment consisting of 
15 servers and in simulations with 50 servers.}
\plb's design, in which the only state clients 
maintain is a probe pool of bounded size, is better 
suited to load balancing at the scale of a system 
such as YouTube.

\resub{Many existing load balancing systems 
(\eg NGINX \cite{nginx-load-balancing}, Envoy \cite{envoy-homepage}, Finagle \cite{finagle-po2c}, 
YARP \cite{YARP-homepage}) 
offer some variant of \podc, 
and most of them offer (client-local) 
RIF, latency, or some combination as the load 
balancing signal. As detailed in \Cref{sec:selection-rule}, 
the use of signals local to the client or load-balancer 
adversely affects performance
at scale, when many clients or balancers might 
be sharing the same pool of replicas.
}

A number of other production systems leverage \bkedit{\podc} load balancing. 
Netflix's Zuul~\cite{Netflix-edge-balancing} incorporates decentralized routing of requests from load 
balancers to backends by randomly considering two candidates and using a combination of local and 
piggybacked RIF statistics from previous responses to choose the least-loaded server. Unlike \plb, 
Zuul eschews active probing and does not incorporate latency information. NGINX's 
implementation~\cite{nginx-load-balancing}, like Zuul, is probe-less, but its 
selection criterion is configurable: ``least-connections'' uses the lowest RIF from a given balancer 
to each backend and ``least-time'' uses estimated latency based on prior requests and the number of 
current connections. 
Given our experience with \plb, we believe it is likely 
that the performance of the load-balancing 
mechanisms described in \cite{Netflix-edge-balancing,nginx-load-balancing} 
could be improved by supplementing the piggybacked data with 
active probing.

\ifblind
\else

\section*{Acknowledgments}

We wish to recognize the contributions of our current and 
former Google colleagues Melika Abolhassani, Doug Rohde, and 
Aliaksei Kandratsenka, who built and experimented extensively 
with an earlier prototype of \plb. Their previous 
yeoman's work gave us a large head start on the current 
iteration of the project. We have also benefitted 
from helpful discussions with David Applegate, Thomas Adamcik, 
Pratik Worah, and Soheil Yeganeh, along with excellent suggestions 
from Philip Fisher-Ogden on the exposition.

\fi

\bibliographystyle{plain}
\bibliography{plb}

\begin{thebibliography}{10}

\bibitem{Kanellakis-Award-2020-power-of-two-choices}
2020 {ACM} {Paris Kanellakis Theory and Practice Award}.
\newblock
  \url{https://www.acm.org/media-center/2021/may/technical-awards-2020}, 2020.
\newblock Accessed: 2023-05-02.

\bibitem{azar99balanced}
Yossi Azar, Andrei~Z. Broder, Anna~R. Karlin, and Eli Upfal.
\newblock Balanced allocations.
\newblock {\em {SIAM} J. Comput.}, 29(1):180--200, 1999.

\bibitem{berenbrink06heavily}
Petra Berenbrink, Artur Czumaj, Angelika Steger, and Berthold V\"{o}cking.
\newblock Balanced allocations: {T}he heavily loaded case.
\newblock {\em SIAM J. Comput.}, 35(6):1350–1385, jun 2006.

\bibitem{Google-SRE-book-load-balancing}
Alejandro~Forero Cuervo.
\newblock Load balancing in the datacenter.
\newblock In Jennifer Petoff, Niall Murphy, Betsy Beyer, and Chris Jones,
  editors, {\em Site Reliability Engineering: How {Google} Runs Production
  Systems}, chapter~20, pages 231--247. {O'Reilly}, Sebastopol, 2016.

\bibitem{delgado2016eagle}
Pamela Delgado, Diego Didona, Florin Dinu, and Willy Zwaenepoel.
\newblock Job-aware scheduling in eagle: {D}ivide and stick to your probes.
\newblock In Marcos~K. Aguilera, Brian Cooper, and Yanlei Diao, editors, {\em
  Proceedings of the Seventh {ACM} Symposium on Cloud Computing, Santa Clara,
  CA, USA, October 5-7, 2016}, pages 497--509. {ACM}, 2016.

\bibitem{envoy-github}
Envoy github repository.
\newblock \url{https://github.com/envoyproxy/envoy}.
\newblock Accessed: 2023-09-21.

\bibitem{envoy-homepage}
Envoy homepage.
\newblock \url{https://www.envoyproxy.io}.
\newblock Accessed: 2023-09-21.

\bibitem{finagle-po2c}
Deterministic {Aperture}: A distributed, load balancing algorithm.
\newblock
  \url{https://blog.twitter.com/engineering/en_us/topics/infrastructure/2019/daperture-load-balancer},
  2019.
\newblock Accessed: 2023-09-21.

\bibitem{gRPC-software-WRR}
{gRPC} {Weighted Round Robin} load balancing component.
\newblock
  \url{https://github.com/grpc/grpc/tree/master/src/core/ext/filters/client_channel/lb_policy/weighted_round_robin}.
\newblock Accessed: 2023-05-03.

\bibitem{Google-SRE-SLO}
Chris Jones, John Wilkes, Niall Murphy, and Cody Smith.
\newblock Service level objectives.
\newblock In Jennifer Petoff, Niall Murphy, Betsy Beyer, and Chris Jones,
  editors, {\em Site Reliability Engineering: How {Google} Runs Production
  Systems}, chapter~4, pages 37--48. {O'Reilly}, Sebastopol, 2016.

\bibitem{katevenis1991weighted}
Manolis Katevenis, Stefanos Sidiropoulos, and Costas Courcoubetis.
\newblock Weighted round-robin cell multiplexing in a general-purpose {ATM}
  switch chip.
\newblock {\em {IEEE} Journal on Selected Areas in Communications},
  9(8):1265--1279, 1991.

\bibitem{kogias2019r2p2}
Marios Kogias, George Prekas, Adrien Ghosn, Jonas Fietz, and Edouard Bugnion.
\newblock $\{$R2P2$\}$: Making $\{$RPCs$\}$ first-class datacenter citizens.
\newblock In {\em 2019 USENIX Annual Technical Conference (USENIX ATC 19)},
  pages 863--880, 2019.

\bibitem{los22balanced}
Dimitrios Los, Thomas Sauerwald, and John Sylvester.
\newblock Balanced allocations: {C}aching and packing, twinning and thinning.
\newblock In {\em Proceedings of the 2022 Annual ACM-SIAM Symposium on Discrete
  Algorithms (SODA)}, pages 1847--1874, 2022.

\bibitem{los23memory}
Dimitrios Los, Thomas Sauerwald, and John Sylvester.
\newblock Balanced allocations with heterogeneous bins: {T}he power of memory.
\newblock In {\em Proceedings of the 2023 Annual ACM-SIAM Symposium on Discrete
  Algorithms (SODA)}, pages 4448--4477, 2023.

\bibitem{mitz00useful}
Michael Mitzenmacher.
\newblock How useful is old information?
\newblock {\em {IEEE} Trans. Parallel Distributed Syst.}, 11(1):6--20, 2000.

\bibitem{mitzenmacher02memory}
Michael Mitzenmacher, Balaji Prabhakar, and Devavrat Shah.
\newblock Load balancing with memory.
\newblock In {\em The 43rd Annual IEEE Symposium on Foundations of Computer
  Science, 2002. Proceedings.}, pages 799--808. IEEE, 2002.

\bibitem{mitz-survey}
Michael Mitzenmacher, Andr\'{e}a~W. Richa, and Ramesh Sitaraman.
\newblock The power of two random choices: {A} survey of techniques and
  results.
\newblock In Sanguthevar Rajasekaran, Panos~M. Pardalos, and Jos\'{e} Rolim,
  editors, {\em Handbook of Randomized Computing}, volume~1, pages 255--312.
  Springer Science \& Business Media, 2001.

\bibitem{Mitzenmacher96-phdthesis}
Michael~David Mitzenmacher.
\newblock {\em The Power of Two Choices in Randomized Load Balancing}.
\newblock PhD thesis, UC Berkeley, 1996.

\bibitem{nginx-load-balancing}
Nginx and the \enquote{Power of Two Choices} load-balancing algorithm.
\newblock
  \url{https://www.nginx.com/blog/nginx-power-of-two-choices-load-balancing-algorithm/},
  2018.
\newblock Accessed: 2023-05-04.

\bibitem{ousterhout2013sparrow}
Kay Ousterhout, Patrick Wendell, Matei Zaharia, and Ion Stoica.
\newblock Sparrow: {D}istributed, low latency scheduling.
\newblock In {\em Proceedings of the 24th ACM Symposium on Operating Systems
  Principles (SOSP)}, pages 69--84, 2013.

\bibitem{park2011generalization}
Gahyun Park.
\newblock A generalization of multiple choice balls-into-bins.
\newblock In {\em Proceedings of the 30th Annual ACM SIGACT-SIGOPS Symposium on
  Principles of Distributed Computing}, pages 297--298, 2011.

\bibitem{Netflix-edge-balancing}
Rethinking {Netflix's} edge load balancing.
\newblock
  \url{https://netflixtechblog.com/netflix-edge-load-balancing-695308b5548c},
  2018.
\newblock Accessed: 2023-05-04.

\bibitem{suresh15c3}
Lalith Suresh, Marco Canini, Stefan Schmid, and Anja Feldmann.
\newblock C3: {C}utting tail latency in cloud data stores via adaptive replica
  selection.
\newblock In {\em 12th USENIX Symposium on Networked Systems Design and
  Implementation (NSDI 2015)}, pages 513--527, 2015.

\bibitem{grpc-blog}
Varun Talwar.
\newblock {gRPC}: a true internet-scale {RPC} framework is now 1.0 and ready
  for production deployments.
\newblock
  \url{https://cloud.google.com/blog/products/gcp/grpc-a-true-internet-scale-rpc-framework-is-now-1-and-ready-for-production-deployments},
  2016.
\newblock Accessed: 2023-09-21.

\bibitem{wieder-survey}
Udi Wieder.
\newblock Hashing, load balancing and multiple choice.
\newblock {\em Foundations and Trends{\textregistered} in Theoretical Computer
  Science}, 12(3--4):275--379, 2017.

\bibitem{YARP-homepage}
{YARP: Yet Another Reverse Proxy}.
\newblock \url{https://microsoft.github.io/reverse-proxy/}.
\newblock Accessed: 2023-09-21.

\bibitem{zhu20racksched}
Hang Zhu, Kostis Kaffes, Zixu Chen, Zhenming Liu, Christos Kozyrakis, Ion
  Stoica, and Xin Jin.
\newblock Racksched: {A} microsecond-scale scheduler for rack-scale computers.
\newblock In {\em 14th {USENIX} Symposium on Operating Systems Design and
  Implementation, {OSDI} 2020}, pages 1225--1240. {USENIX} Association, 2020.

\end{thebibliography}
\appendix
\newpage

\section{\resub{Linear combinations of latency and RIF}}
\label{sec:lincombo}

\begin{figure}[htb]
\centering
\begin{tikzpicture}
    \draw (-0.05,5.5) node {\includegraphics[width=0.515\textwidth]{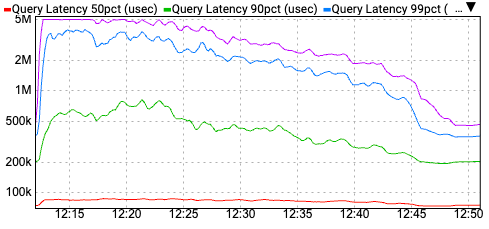} };
    \draw (0,0) node {\includegraphics[width=0.505\textwidth]{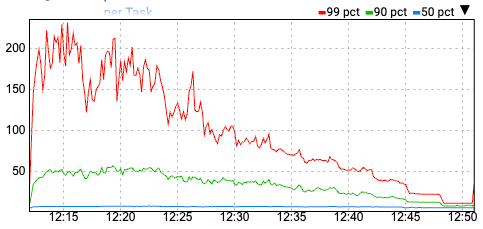} };
    \def\xbase{-3.95}
    \def\ybase{-1.8}
    \def\ytop{8.8}
    \def\topbarht{1}
    \def\sw{0.638}
    \def\coeff{{.769,.785,.801,.817,.834,.868,.886,.904,.922,.941,.960,.980,1.0}}
    \foreach \i in {0,1,2,3,4,5,6} {
        \fill[ultra nearly transparent] (\xbase + 2*\sw*\i,\ybase) rectangle (\xbase + \sw + 2*\sw*\i,\ytop);
        \tikzmath{\stripelabel = \coeff[2*\i];}
        \draw (\xbase + 0.5*\sw + 2*\sw*\i,\ytop - 0.8) node {\scriptsize \stripelabel};
        \ifnum \i < 6 
        \fill[ultra nearly transparent] (\xbase + \sw + 2*\sw*\i,\ytop - \topbarht) rectangle (\xbase + 2*\sw + 2*\sw*\i,\ytop);
        \tikzmath{\stripelabel = \coeff[1 + 2*\i];}
        \draw (\xbase + 1.5*\sw + 2*\sw*\i,\ytop - 0.8) node {\scriptsize \stripelabel};
        \fi
    }
    \draw (0,\ytop - 0.3) node {\bf Coefficient of RIF};
    \draw (0,3) node {(a) Latency quantiles};
    \draw (0,-2.5) node {(b) RIF quantiles};
\end{tikzpicture}
\caption{Evaluating replica selection rules based on 
linear combination of latency
and RIF.}
\label{fig:lincombo}
\end{figure}
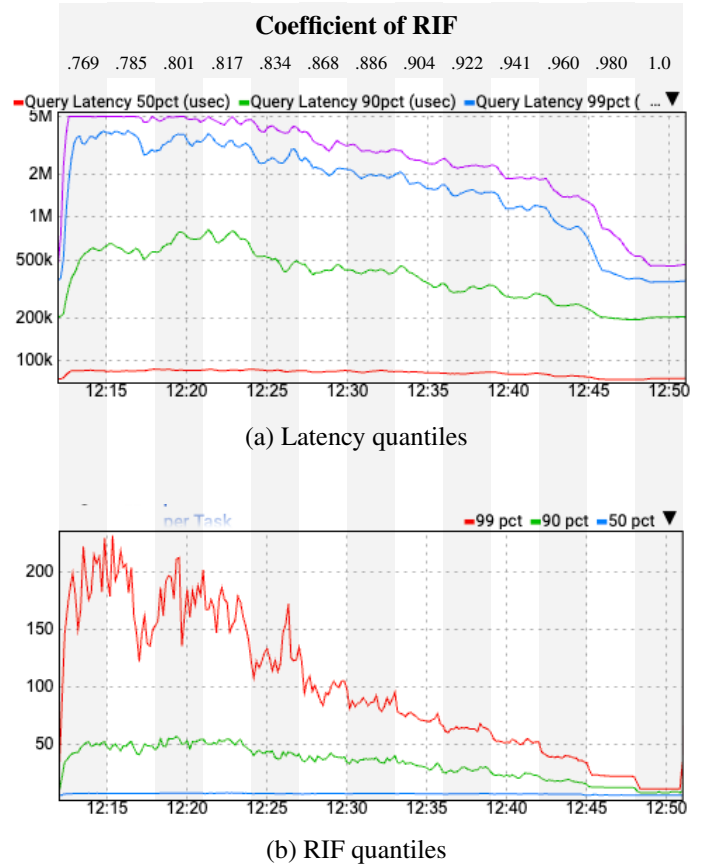
\resub{%
To evaluate the effectiveness of replica selection rules that use a linear combination of latency and RIF, we used our testbed\footnote{See the start of \Cref{sec:evaluation} for a general description of the testbed environment.}
to experiment with a variant of \plb that uses the some asynchronous probing method detailed in \Cref{sec:design}, modified to select replicas using a linear combination of latency and RIF. In other words, the \hotcold replica selection rule was replaced with one that chooses among the replicas represented in the probe pool by minimizing the score
\begin{equation}
    \label{eq:lincombo}
    \text{score}^{\lambda}_i = 
    (1 - \lambda) \cdot \text{latency}_i + 
    \lambda \cdot \alpha \cdot \text{RIF}_i
\end{equation}
where $\text{latency}_i, \, \text{RIF}_i$, respectively, 
denote the latency and RIF in the probe response from 
replica $i$, $\alpha$ is a scale factor applied to 
RIF to convert it into the same units as latency, and
$\lambda \in [0,1]$ is a tunable parameter that adjusts
the relative weight given to latency and RIF. Setting
$\lambda = 0$ corresponds to latency-only control, whereas
$\lambda = 1$ corresponds to RIF-only control.

To set the scale factor $\alpha$, we used the
approximate median query response time for server 
replicas with one request in flight. This value 
turned out to be 75 milliseconds. Note that for
any $\alpha > 0$, the 
set of scoring rules $\{ \text{score}^{\lambda}
\mid 0 \le \lambda \le 1 \}$ obtained by varying
$\lambda$ over the interval $[0,1]$ is always
equal to the set of convex combinations of latency
and RIF. In other words our choice of 
$\alpha = 75\text{ms}$
affects the way that this set of scoring rules is
parameterized by $\lambda$, but doesn't affect the
set itself. 

We evaluated linear combination scoring rules by measuring quantiles of latency and RIF in our testbed at a constant level of aggregate CPU utilization (equal to 94\% of our allocation) while varying $\lambda$.
\bkedit{Replicas were partitioned into an equal number of fast (odd-numbered) and slow (even-numbered) ones, with a 2x difference in query processing speed, as in the RIF quantile experiment described in \Cref{sec:tunable}.}
We initially experimented with varying $\lambda$
over the full range $[0,1]$ in increments of 0.1.
It was evident from this initial experiment that
linear combination rules with $\lambda \le 0.7$
performed poorly compared to larger values of 
$\lambda$, which prompted us to examine the range
depicted in \Cref{fig:lincombo} at a finer resolution. 
Of the 13 linear combinations tested in the experiment, all quantiles of latency and RIF improved monotonically as $\lambda$ increased, with $\lambda=1$ (i.e., RIF-only control) dominating all other linear-combination feedback control rules, in most cases by a wide margin.

Recall from \Cref{fig:rif-limit} in \Cref{sec:tunable} that RIF-only control (represented in that figure by RIF limit threshold 1.0, the rightmost configuration) performs strictly {\em worse} than \plb at all quantiles of latency and RIF. Since the results of the experiment reported here show that RIF-only control performs strictly {\em better} than any other linear combination of latency and RIF, it follows by transitivity that \plb strictly dominates all linear combinations of latency and RIF. \bkdelete{Accordingly, in the experiment on replica selection reported in \Cref{sec:selection-rule}, we chose to use just one linear combination of latency and RIF (corresponding to $\lambda=0.5$) as a stand-in for the entire parameter space of linear combinations. }
}

\end{document}